\documentclass[]{aastex631}
\usepackage{appendix}
\usepackage{graphicx}
\usepackage{comment}
\usepackage{float}

\usepackage{lineno} % 添加此行以启用行号包
\begin{document}

% \title{CURLING – III. Searching for Galaxy Cluster-Scale Strong lenses from the DESI Legacy Imaging Surveys}

\title{Searching for Galaxy Cluster-Scale Strong lenses from the DESI Legacy Imaging Surveys}

% \thanks{zhejian-23@mails.tsinghua.edu.cn}
\author{Zhejian Zhang}
\affiliation{Department of Astronomy, Tsinghua University, Beijing 100084, China}

\thanks{nan.li@nao.cas.cn}
\author{Nan Li}
\affiliation{National Astronomical Observatories, Chinese Academy of Sciences, 20A Datun Road, Chaoyang District, Beijing 100012, China 
}%通讯作者

\author{Shude Mao}
\affiliation{Department of Astronomy, School of Science, Westlake University, Hangzhou, Zhejiang 310030, China}

\author{Hu Zou}
\affiliation{National Astronomical Observatories, Chinese Academy of Sciences, 20A Datun Road, Chaoyang District, Beijing 100012, China 
}

\author{Zizhao He}
\affiliation{Department of Physics, Nanchang University, Nanchang, 330031, China}
\affiliation{Center for Relativistic Astrophysics and High Energy Physics, Nanchang University, Nanchang, 330031, China}
\affiliation{Purple Mountain Observatory, Chinese Academy of Sciences, Nanjing, Jiangsu, 210023, China}

\author{Mingxiang Fu}
\affiliation{School of Astronomy and Space Science, University of Chinese Academy of Sciences, Beijing 101408, China}
\affiliation{National Astronomical Observatories, Chinese Academy of Sciences,Beijing 100101, China}

\author{Shenzhe Cui}
\affiliation{School of Astronomy and Space Science, University of Chinese Academy of Sciences, Beijing 101408, China}
\affiliation{National Astronomical Observatories, Chinese Academy of Sciences,Beijing 100101, China}

\begin{abstract}

Galaxy cluster-scale strong gravitational lensing systems are rare yet valuable tools for investigating the properties of dark matter and dark energy, as well as providing the opportunity to study the distant universe at flux levels and spatial resolutions that would otherwise be unavailable. Large-scale imaging surveys present unprecedented opportunities to expand the sample of cluster lenses. In this study, we adopt a deep learning-based approach to identify cluster lenses from the DESI Legacy Imaging Surveys, utilizing the catalog of galaxy cluster candidates identified by \cite{Zou_2021}. Our lens-finder employs a ResNet-18 architecture, trained with mock images of cluster lenses as positives and observational images of cluster scale non-lenses as negatives. We do an iterative operation to increase the completeness of our work, namely adding the found true positive samples back to the training set and training again for several times. Human inspection is conducted to further refine the candidates, categorizing them into grades (A, B, C) according to the significance of the strongly lensed arcs. Reviewing all 540,432 objects in Zou's catalog, we discover 485 high-confidence cluster lens candidates with a cluster $M_{500}$ range of $10^{13.67\sim14.97}M_{\odot}$ and a Brightest Central Galaxy (BCG) redshift range of $0.04\sim0.89$. After excluding the lens candidates listed in previous studies, we identify 247 newly discovered cluster lens candidates, including 16 grade A, 90 grade B, and 141 grade C. This catalog of cluster lens candidates is publicly available online{\footnote{\url{https://github.com/zkzdyzg/Galaxy-Cluster-Scale-Strong-lenses-from-the-DESI-Legacy-Imaging-Surveys}}}, and follow-up observations are encouraged to confirm and conduct thorough investigations of these systems.
\end{abstract}

\keywords{Strong Gravitational Lenses - Galaxy Cluster-  Convolutional Neural Network}

\section{Introduction} \label{Sec: Intro}

Strong gravitational lensing (SGL) refers to the noticeable distortion of a distant source's image caused by the gravitational field of massive objects located between the observer and the source \citep{1996astro.ph..6001N}. This phenomenon produces distinctive features, such as Einstein rings, arcs, or multiple images. Strong gravitational lensing systems serve as a valuable tool for cosmological research, enabling the study of the distribution of mass, particularly dark matter, within lensing objects (e.g. \citealt{1988MNRAS.231P..97N,2010ARA&A..48...87T,2010RPPh...73h6901M, 2015ApJ...800...38G,2020Sci...369.1347M}). Time-delay measurements between multiple lensed images provide a geometric method to constrain the Hubble constant (e.g. \citealt{1964MNRAS.128..307R,1996ApJ...464...92G, 2017MNRAS.468.2590S,2020MNRAS.498.1420W}). Additionally, the magnification effect of SGL facilitates the observation of intrinsically faint populations at high redshifts, offering insights into early star formation and galaxy assembly (e.g. \citealt{2018NatAs...2..334K,2022Natur.603..815W,2024A&A...687A..81P}).

Cluster-scale strong lensing systems, where massive galaxy clusters ($M_{500}\gtrsim 10^{14}M_{\odot}$) act as gravitational lenses, produce distinctive arcs or ring structures, offering numerous scientific opportunities (e.g. \citealt{1984Natur.310..112N,1996ApJ...471..643K,2011A&ARv..19...47K,2007NJPh....9..447J,2010MNRAS.404..325R,2012ApJS..199...25P,2019ApJ...884...85C}). These systems provide valuable insights into both astrophysical and cosmological aspects. Their sensitivity to gravitational effects enables the mapping the cluster mass distributions \citep{1993A&A...273..367K,2009MNRAS.395.1319J,2012ApJ...757...22C,2015MNRAS.452.1437J}, constraining the properties of dark matter or cosmology \citep{2010Sci...329..924J,2016A&A...587A..80C}, and probing the mechanisms of cluster formation \citep{2015ApJ...806....4M}. Moreover, the magnification of high redshift background sources enables the study of galaxy formation during the earliest epochs, which extended to the dark age \citep{2004ApJ...607..697K,2013ApJ...762...32C}. Furthermore, previous studies demonstrate that light deflection information can constrain cosmological parameters, offering a powerful geometric tool to probe Dark Energy \citep{2010Sci...329..924J}. 

Galaxy cluster-scale strong lensing systems are, however, rare phenomena \citep{2008ApJS..176...19F,2010PASJ...62.1017O}. The advent of wide-field sky surveys, such as the Dark Energy Spectroscopic Instrument (DESI) Legacy Imaging Surveys \citep{2019AJ....157..168D}, the Vera C. Rubin Observatory’s Legacy Survey of Space and Time (LSST) \citep{2019ApJ...873..111I}, and Euclid \citep{2025arXiv250315324E}, has increased the opportunity to discover these rare systems. With the vast amount of data generated, many strong lensing events likely remain undiscovered. However, identifying them is a challenge \citep{2019A&A...625A.119M}. Traditional detection pipelines, which rely on visual inspection \citep{2021ApJ...906..107K} or parametric morphology filters (e.g., arc length and curvature thresholds) \citep{2004A&A...416..391L,2005ApJ...633..768H,2006astro.ph..6757A,2007A&A...472..341S,2016ApJ...817...85X}, scale poorly to such massive data volumes. Manual vetting becomes prohibitively time-intensive, whereas rule-based algorithms suffer from high false-positive rates due to contaminants such as spiral arms, tidal debris, and edge-on galaxies. 

Recent advancements in machine learning, particularly in Convolutional Neural Networks (CNNs), have shown remarkable potential for detecting strong gravitational lenses in astronomical imaging data. Trained on large-scale simulations of lensing systems, CNNs are highly effective at identifying subtle morphological features, such as faint arcs or distorted counterimages, which often evade traditional detection algorithms \citep{2019ApJS..243...17J}. Early applications of CNN-based classifiers, such as those employed in the Hyper Suprime-Cam Subaru Strategic Program (HSC-SSP) \citep{2024MNRAS.535.1625J}, the Kilo-Degree Survey (KiDS) \citep{2017MNRAS.472.1129P,2021ApJ...923...16L} and the Dark Energy Spectroscopic Instrument (DESI) Legacy Surveys \citep{Huang_2020,2021ApJ...909...27H, 2024ApJS..274...16S}, have achieved recall rates of $\geq90\%$ for galaxy-scale lenses. These methods reduce false positive rates by an order of magnitude compared to rule-based approaches, demonstrating their effectiveness in addressing the challenges of lens discovery in massive datasets.

In this study, we present a CNN-based methodology designed to identify cluster-scale strong gravitational lensing systems within the cluster catalog compiled by \cite{Zou_2021} from the DESI Legacy Imaging Surveys. Our method employs an 18-layer ResNet architecture, chosen for its demonstrated effectiveness in extracting features from complex astronomical images, to systematically detect strong lensing signatures within galaxy clusters. To train the model, we generate realistic cluster-scale lensing images using a mock program that combines DESI legacy imaging of real galaxy clusters with synthetic lensed background sources, incorporating variations in mass distribution, redshift, and observational conditions. The initial CNN model, trained on mocked data, is applied to the DESI fields to identify strong lensing candidates. The confirmed true positives, obtained through visual inspection, are subsequently incorporated into the training set for iterative retraining of the model. our pipeline successfully identified 485 cluster-scale strong lensing candidates, where 247 of them are newly found compared to other work in DESI data.

This article is organized as follows: In Section \ref{Sec: data}, we introduce the data we used, namely the galaxy cluster catalog from \cite{Zou_2021}, and then describe our mock program used to produce mock lensing images to generate our training set and validation set. Section \ref{Sec: method} provides the construction of our neural network model and the training process, then in Section \ref{Sec: result} we present the strong lensing candidates we found. Finally in Section \ref{Sec: discussion}, we discuss our findings and draw a conclusion. Throughout this paper, all RGB color composite images are generated using the $g$, $r$, and $z$ band images from the DESI Legacy Imaging Surveys, which are assigned to the blue, green, and red color channels, respectively.

\section{Data}\label{Sec: data}

In this section, we first describe the 540,432 galaxy clusters identified in the DESI Legacy Imaging Surveys, as reported by \cite{Zou_2021}, in Subsection \ref{subsec: obser}. These clusters may serve as input for our CNN-based methodology to identify strong gravitational lenses. Next, we provide details on the mock framework developed to generate simulated lens images for training in Subsection \ref{subsec: mock}, along with the preprocessing steps applied to the data in Subsection \ref{subsec: process}.

\subsection{Observation data}\label{subsec: obser}

The DESI Legacy Imaging Surveys (DESI-LIS) integrate three optical surveys to support spectroscopic target selection for the Dark Energy Spectroscopic Instrument (DESI): BASS (the Beijing-Arizona Sky Survey, \citealt{2017PASP..129f4101Z}), DECaLS (the Dark Energy Camera Legacy Survey, \citealt{2016AAS...22831701B}) and MzLS (the Mayall $z$-band legacy survey, \citealt{2016AAS...22831702S}) three surveys, which covered a sky area of $\sim14,000 \ \rm deg^2$ in the $g, r,$ and $z$ bands, and also integrated the latest WISE observations in the W1 and W2 bands \citep{2016AJ....151...36L, 2019PASP..131l4504M}. The cluster catalog used in this work is based on the latest data released (DR8) from DESI legacy imaging surveys at the time \cite{Zou_2021} was published, which covered an area of approximately $20,000 \  \rm deg^{2}$. Details regarding the galaxy cluster images used in this study can be found in \cite{Zou_2021}.

These clusters were identified based on the photometric redshifts derived from DESI galaxies. As discussed in the work of \cite{Zou_2019}, they used the five-band deep photometry of the DESI legacy imaging surveys to obtain a catalog of accurate photometric redshifts and stellar masses for classified galaxies. They also applied some quality cuts to eliminate galaxies with very large photo-z uncertainty or very low luminosity. Based on the redshift catalog, \cite{Zou_2021} applied the CFSFDP approach \citep{2014Sci...344.1492R} similar to \cite{2020PASP..132b4101G} to detect clusters. A total of 540,432 galaxy clusters at $z \lesssim 1$ were finally identified. See Figure \ref{fig:result_pos} for the number density of these clusters. The quality of this catalog is high enough to support our subsequent work. Monte Carlo simulation was performed to evaluate their detection approach and estimate the false detection rate, which showed that the false detection rate is about 3.1\%, and the false rate can reach up to about 8\%. The mass and richness of these clusters were also estimated by X-ray and radio observations, which were approximately $1.23\times 10^{14} M_{\odot}$ and 22.5 respectively. %The cluster catalog covers almost all Abell clusters \citep{1989ApJS...70....1A} and most of the clusters identified by SDSS data (e.g. \citealt{2007ApJ...660..239K,2010ApJS..191..254H,2014ApJ...785..104R,2015ApJ...807..178W}).

\subsection{Mock data}\label{subsec: mock}

The limited number ($<1000$) of confirmed galaxy cluster-scale strong gravitational lenses presents significant challenges for training robust machine learning detectors. Recent galaxy-scale lens catalogs include more than 3,000 candidates \citep{2024ApJS..274...16S}, the expected number of cluster-scale lenses remains approximately 1 order of magnitude lower due to the sparse spatial density of massive halos. 

To address this limitation, a widely used approach is to generate mock lensing images. We develop a physically motivated simulation framework to produce mock cluster-scale lenses, incorporating several key components: (1) mass profiles constrained by observational data, derived from DESI photometric redshifts and luminosity-mass scaling relations; (2) simulated source galaxy populations based on \cite{2015ApJ...811...20C}; (3) wavelength-dependent PSF convolution using DECam instrument profiles; and (4) noise injection calibrated to DESI-LIS depth maps. Specifically, we construct the lensing potential, place a background source from the source catalog, and simulate the resulting arc. After applying PSF convolution and injecting noise, the arc is incorporated into the image.

To enhance realism and expand the parameter space, we also introduce adjustments to the mock lenses, ensuring that they better capture the diversity of cluster-scale lensing systems. This approach significantly improves the utility of our simulations for training machine learning models.

\begin{figure}[htb]
        \centering
        \includegraphics[scale=0.28]{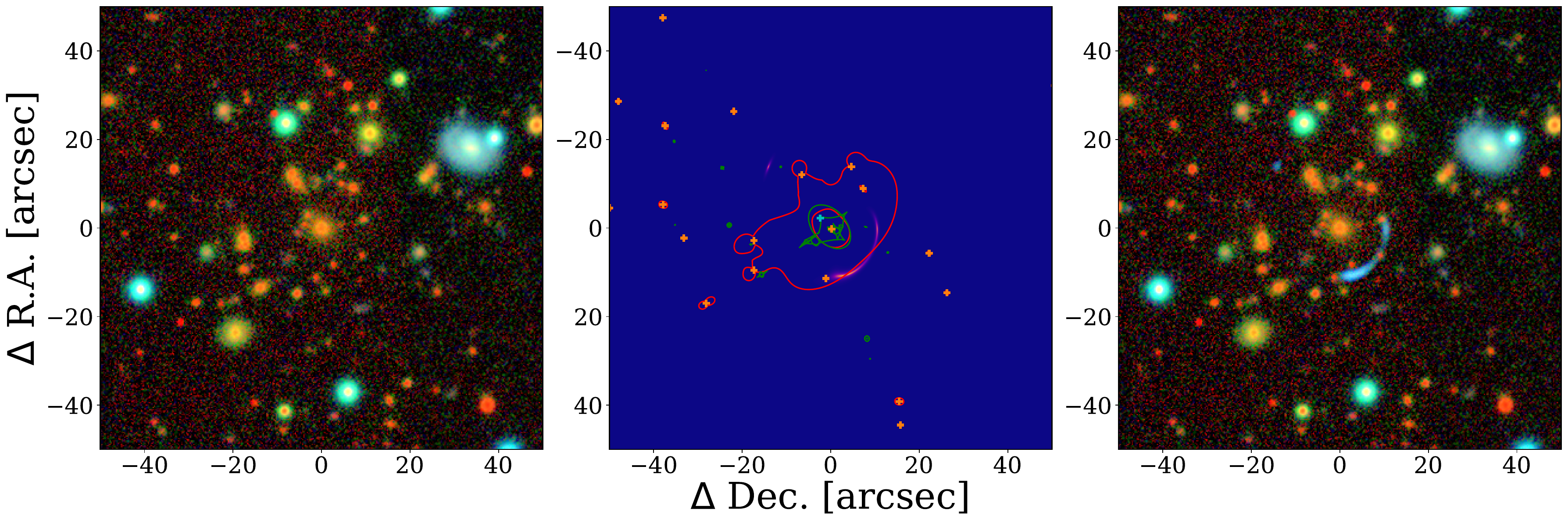}
        
        \caption{One example for our mock program, with side length of $100''$. Left: One selected non-lensed cluster image. Middle: The mock arc image to be added to the cluster image, together with the critical curves (red curve) and the caustics (green curve). The orange crosses indicate the positions of the member galaxies within the galaxy cluster, while the cyan cross indicate the source position.  Right: The final mock lensing images.}
        \label{mock example}
    \end{figure}

The methodology of our mock program is outlined as follows:
\begin{enumerate}
\item Parameter Extraction: We begin by selecting non-lensing clusters from the catalog by \cite{Zou_2021}, obtaining key parameters such as the RA and DEC of the Brightest Central Galaxy (BCG), the mass $M_{500}$, and the characteristic radius $R_{500}$ for each galaxy cluster. Here, the subscript 500 indicates that the average density within $R_{500}$ is 500 times the critical density of the universe, while $M_{500}$ represents the total mass of the galaxy cluster contained within $R_{500}$. Then we extract photometric redshifts, axis ratios, position angles, ellipticities, and stellar mass estimates for each member galaxy from the galaxy cluster member catalog compiled by \cite{Zou_2021}.  
\item Modeling the Lensing Potential: We model all member galaxies within the clusters as Singular-Isothermal-Ellipsoids (SIEs) using the parameters obtained in Step 1. The velocity dispersions are estimated following the method described in Eq. (2) of \cite{2022AJ....163..139Y}. The lensing potential of the galaxy cluster is calculated based on its mass and radius using an elliptical Navarro-Frenk-White (eNFW) profile \citep{1996ApJ...462..563N, 2002A&A...390..821G}. The total lensing potential is determined by combining by summing the NFW potential of the cluster and the SIE potentials of the member galaxies. This combined potential allows us to compute key lensing properties, including deflection angles, magnification maps, critical curves, and caustics. Since the redshifts of the member galaxies are approximately the same, it is reasonable to approximate the total potential by directly summing the contributions from the cluster and its member galaxies.
\item  We adopt a Sérsic profile for the source, with source parameters drawn from the catalog developed by \cite{2015ApJ...811...20C}, which provides a population of statistically realistic galaxy–galaxy strong lenses simulated for the Dark Energy Survey (DES) along with their observed properties. From this catalog, we extract key parameters of the sources, including their effective radii ($r_e$), magnitudes in three bands ($g,r,z$) and axis ratios ($q$). Note that we do not adopt the Sérsic index of the source galaxies in \cite{2015ApJ...811...20C}'s catalog, since those values are all set to be unity. In practice, this assumption yields overly long arcs and fails to reproduce multiple images, and it also differs from the Sérsic indices of high-redshift galaxies in the simulation. Instead, we use the Sérsic index distribution of galaxies with redshifts of 1-2 in the simulation \citep{Xu_2019} as the Sérsic index distribution of our source galaxies. After defining the source properties, we place the centers of the mock sources in regions with high magnifications to generate lensed images.
\item Selection criteria for realistic arcs: We apply a selection criterion based on the size and color of the arcs to ensure that our training set spans a broad parameter space. Galaxy clusters with Einstein radius smaller than $5''$ are excluded from the final sample. The galaxies in the source catalog are predominantly high-redshift blue galaxies. However, during preliminary investigations, we identified several strong lensing systems with red arcs, which are underrepresented represented in the original catalog. To address this, we define three color categories based on the $g - r$ color index: red ($r+1.3 \leq g$), green ($r+0.8 \leq g \leq r+1.3$), and blue ($g \leq r+0.8$). The ratio of sources in these categories is set to 1:1:2, respectively, to better reflect the diversity of colors observed in real lensing systems. 
\item Incorporating Point-Spread function (PSF) and noise: We convolve with a Gaussian PSF and then add Poisson noise. The PSF parameters are consistent with those of real PSF in the DESI Legacy Imaging Survey. Finally, we overlay the resulting arcs onto the original cluster images to obtain the final mock lenses. 
\end{enumerate}

\begin{table}
	\centering
	\caption{Range and distribution of parameters used to generate the mock lensed images. Distributions for parameters in this table without analytic forms are shown in Figure \ref{fig:mock_parameter}.}
	\label{tab:mock_parameter}
	\begin{tabular}{llll} % four columns, alignment for each
		\hline
        \hline
		Parameter & Range & Distribution\\
		\hline
        \hline
            lens (eNFW)\\
        \hline
        $\log_{10}(M_{500}/M_{\odot})$ & $12.58 \sim 15.13$ & - \\
        Photometric redshift (BCG) & 0.01 $\sim$ 1.04 & - \\
		Position angle & $-90^{\circ}$ $\sim$ $-90^{\circ}$ & uniform\\
		Axis ratio & 0.4 $\sim$ 1.0  & normal($\mu=0.75, \sigma=0.1$)\\
		\hline
            lens (SIE members)\\
        \hline
        $\log_{10}(M_{*}/M_{\odot})$ & $8.04 \sim 12.69$  & - \\
		% Position angle & -90 $\sim$ 90  & degree & -\\
		% Axis ratio & 0.009 - 1.0  & - & -\\
        \hline
            source (Sérsic)\\
        \hline
        $g$-band magnitude & 21.3 $\sim$ 29.5 & -\\
        $r$-band magnitude & 21.0 $\sim$ 28.0 & -\\
        $z$-band magnitude & 20.3 $\sim$ 28.3 & -\\
        Effective radius & $0.00''$ $\sim$ $1.68''$ & -\\
        axis ratio & 0.2 $\sim$ 1.0 & -\\
        Redshift & 0.034 $\sim$ 1.495 & -\\
        Sérsic index & 0.4 $\sim$ 4.0 & -\\
	\end{tabular}
\end{table}

\begin{figure}
    \centering
    \includegraphics[width=0.9\linewidth]{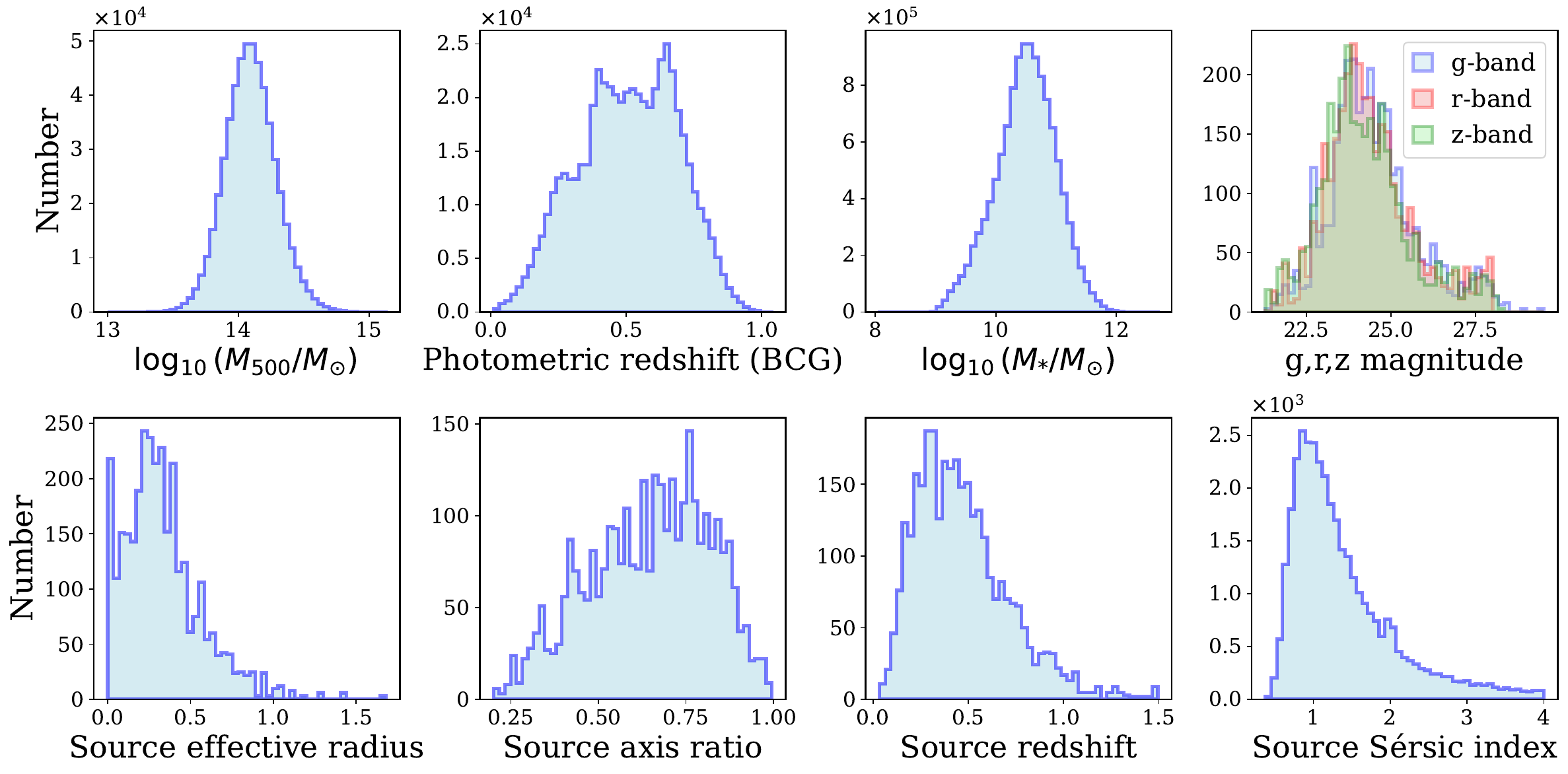}
    \caption{Distributions for parameters without analytic forms in Table \ref{tab:mock_parameter}. The source effective radius is units of arcseconds.} 
    \label{fig:mock_parameter}
\end{figure}

\begin{figure}
    \centering
    \includegraphics[width=0.45\linewidth]{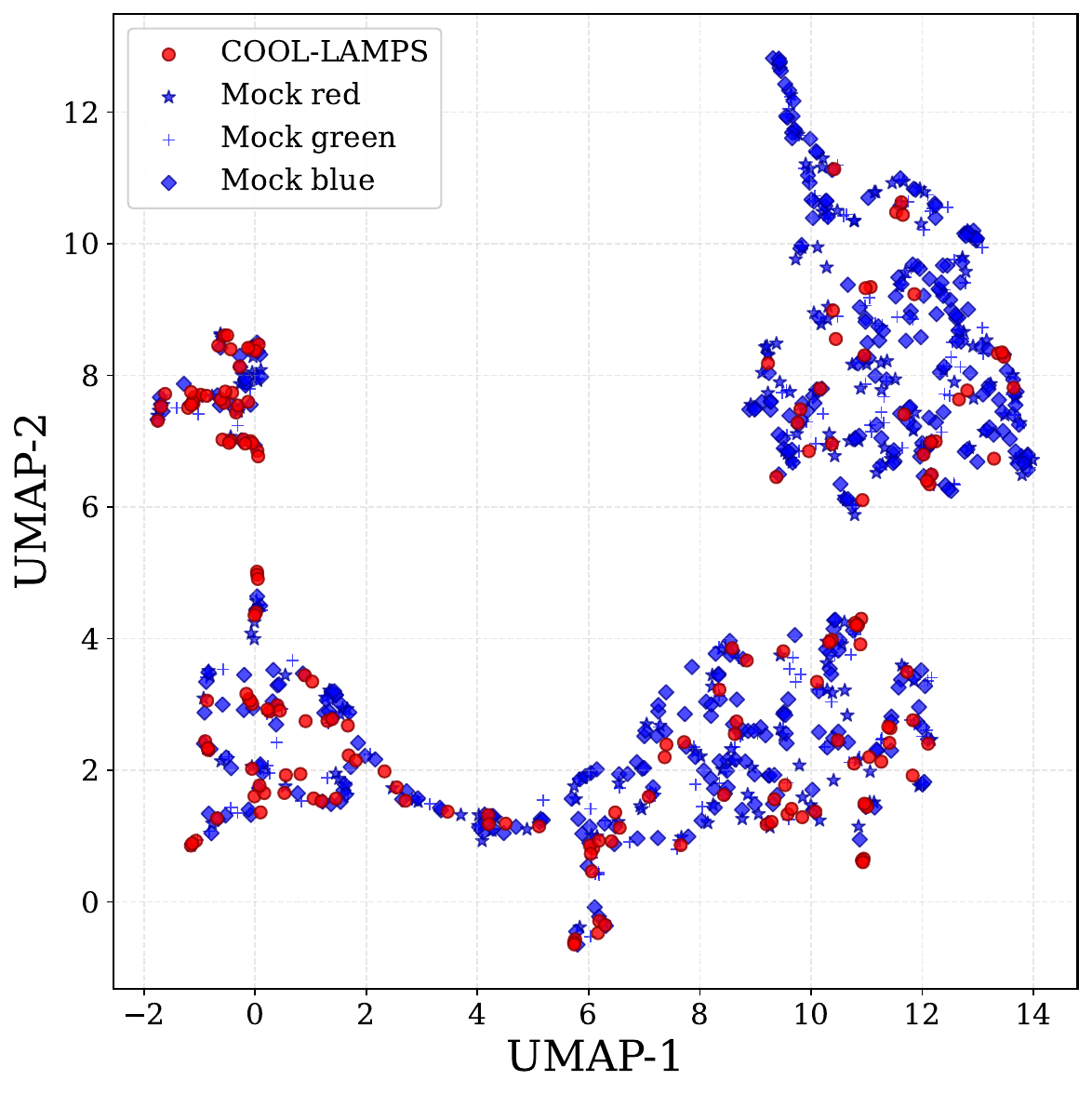}
    \caption{UMAP projection of randomly selected 800 mock lenses from the training set, including the systems with red (blue star), green (blue cross), and blue (blue diamond) sources defined in our manuscript (See Section \ref{subsec: mock}). Red dots present the UMAP of 177 cluster-scale strong lenses from the COOL-LAMPS project \citep{2025ApJ...979..184M}.} 
    \label{fig:UMAP}
\end{figure}

Figure \ref{mock example} shows an example of our mock image, and Table \ref{tab:mock_parameter} gives the range and distribution of the parameters in our simulation. To assess the similarity between our mock images and real observations, we perform an unsupervised clustering analysis using the Uniform Manifold Approximation and Projection (UMAP; \citealt{mcinnes2020umapuniformmanifoldapproximation}) algorithm, which is applied to feature representations extracted from both our mock lenses and the high-quality candidates of cluster-scale strong lenses given by the COOL-LAMPS project (see also subsection 4.2; \citealt{2025ApJ...979..184M}). The resulting UMAP visualization in Figure \ref{fig:UMAP} shows that the mock lenses and the COOL-LAMPS lenses occupy overlapping regions in the reduced-dimensional feature space, suggesting that our mock images do not exhibit significant systematic differences from the COOL-LAMPS lenses in the feature space captured by the UMAP algorithm, thus supporting the fidelity and applicability of the mocks for training strong lens classifiers based on CNNs.

\subsection{Data preprocessing}\label{subsec: process}

Using our framework, we first select a sufficient number of non-lensed cluster images. These images are obtained in the photometric bands $g$, $r$, and $z$, with a default pixel size of $0.27''$ and dimensions of $371\times371$ pixels, corresponding to an approximate angular size of $100''$. Larger images are downloaded to accommodate potential subsequent tasks, such as cropping, thereby enhancing the flexibility of our analysis. Half of these images are used to generate mock lenses and are treated as positive samples (P), while the remaining half are left unprocessed and treated as negative samples (N).

To compose the training and validation sets, all images are cropped to a side length of 223 pixels ($60''\times 60''$). The smaller image size allows CNN to better capture the spatial structure of the lens. The final training set and validation set consist of 100,000 images (50,000 P + 50,000 N) and 10,000 images (5,000 P + 5,000 N), respectively.

\section{Methodology}\label{Sec: method}

In this section, we present the methodology of our work. First, we describe the convolutional neural network (CNN) employed and the associated training configuration. Next, we outline the methods employed to evaluate network training. Finally, we describe the iterative process adopted to enhance the results.

Our lens detection system utilizes a modified 18-layer Residual Neural Network (ResNet-18) architecture \citep{he2015deepresiduallearningimage}, reimplemented in PyTorch 1.10 with specific adaptations for processing astronomical images. This adaptation follows a similar approach to that used in \cite{2021ApJ...923...16L}, which has demonstrated efficient performance in comparable applications.  The baseline model is reconfigured to accept 3-channel 223×223 pixel inputs, which correspond to the dimensions of our pre-processed galaxy cluster cut-outs that combine $g$, $r$, and $z$-band observations through chromatic stacking. The architecture consists of five residual block groups that progressively reduce spatial dimensions from 112 × 112, following the initial convolution and max-pooling, to 7 × 7 feature maps. Strided convolutions with a stride of 2 are applied at strategic layers to preserve morphological features critical for lens identification. Each residual block contains two 3×3 convolutional layers with Batch Normalization \citep{2015arXiv150203167I} and ReLU activation \citep{nair2010rectified}, maintaining gradient flow through identity shortcut connections. The network concludes with global average pooling, followed by a 2-dimensional fully connected layer that outputs logits optimized for binary classification between lensing systems and astrophysical false positives.

The network is trained using stochastic gradient descent with the Adam optimizer \citep{kingma2017adammethodstochasticoptimization}, configured with a batch size of $32$, a learning rate of $\eta=3\times10^{-6}$, and a weight decay of $\lambda=10^{-1}$. The cross-entropy loss function \citep{goodfellow2016deep} is minimized:
\begin{equation}
    -\sum_{i=1}^Ny_i\log\hat{y_i}+(1-y_i)\log(1-\hat{y_i}), 
\end{equation}
where $y_i\in \{0,1\}$ is label for the $i$th image (1 for lens and 0 for non-lens), and $\hat{y_i} \in [0, 1]$ is the probability predicted by the network. Training is performed on a Nvidia L40S GPU and required 4 hours to complete 200 epochs.

\begin{figure}
    \centering

    \includegraphics[width=0.9\linewidth]{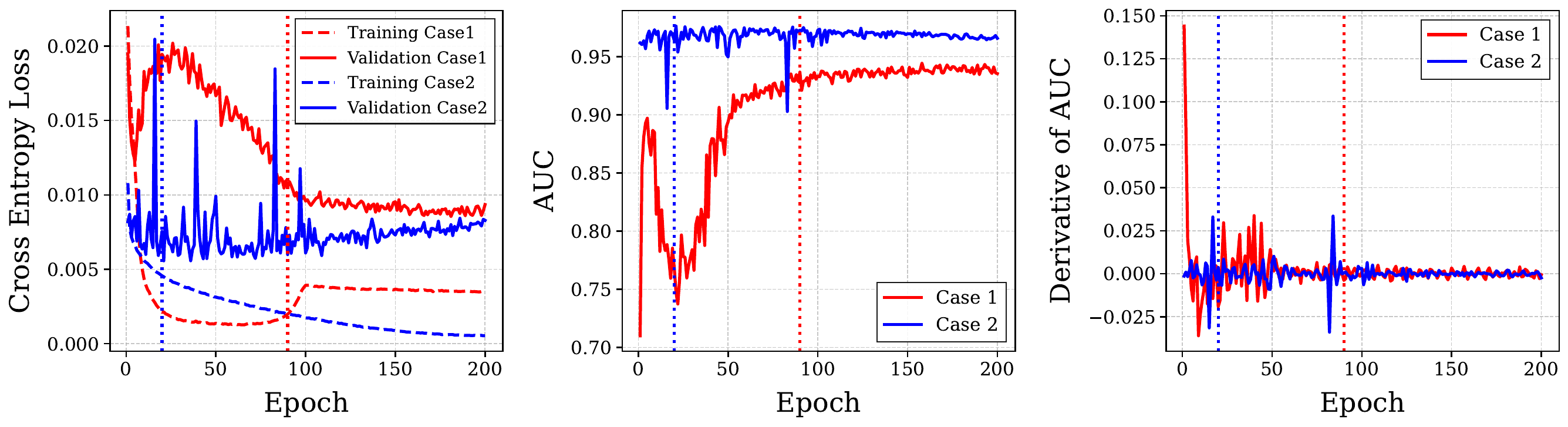}
    \caption{Left: The cross-entropy loss on the training and validation sets over 200 training epochs with two cases in the final iteration of our training. In Case 1, we adopt Adam optimizer, configured with a batch size of 32, a learning rate of $3\times10^{-6}$, and a weight decay of $1\times10^{-1}$. Moreover, in Case 2, We use a initial learning rate of $1\times10^{-4}$, a final learning rate of $1\times10^{-5}$, a weight decay of $1\times10^{-2}$ , with Cosine Annealing. The vertical dashed line at epoch 20 and 90 indicates the early stopping points we selected in the two cases, which is the highest point of the AUC (Case 2), or the highest AUC before the training loss become abnormal (Case 1). Middle: The Area Under the ROC curve as a function of epoch in the same training session. Right: The derivative of AUC as a function of epoch in the same training session.}
    \label{fig:loss}
\end{figure}

We quantify the detection capability of our gravitational lens identification system using Receiver Operating Characteristic (ROC) analysis. The network assigns a lens probability score $\hat{p}\in [0,1]$ to each input image, representing the confidence level of lens detection. By systematically varying the classification threshold $\tau$ across the full probabilistic range $(0 \leq \tau \leq 1)$, samples are dynamically classified as positive ($p\geq\tau$) or negative ($p<\tau$), generating four classification categories: True Positives (TPs, correctly identified lenses), False Positives (FPs, non-lens objects misclassified as lenses), True Negatives (TNs, correctly rejected non-lenses) and False Negatives (FNs, undetected lens systems). The diagnostic metrics are mathematically defined as follows:
\begin{equation}
	\text{TPR} = \frac{\text{TP}}{\text{TP} + \text{FN}},
	\end{equation}
	\begin{equation}
	 \text{FPR} = \frac{\text{FP}}{\text{FP} + \text{TN}},
	\end{equation}
where the True Positive Rate (TPR) measures the completeness of lens recovery, and the False Positive Rate (FPR) quantifies contamination from spurious detections. The ROC curve plots TPR against FPR parametrically across all possible $\tau$ values, with the Area Under the Curve (AUC) serving as a threshold-independent performance metric. 

\begin{figure}[htb]

            \centering
            \includegraphics[width=1\linewidth]{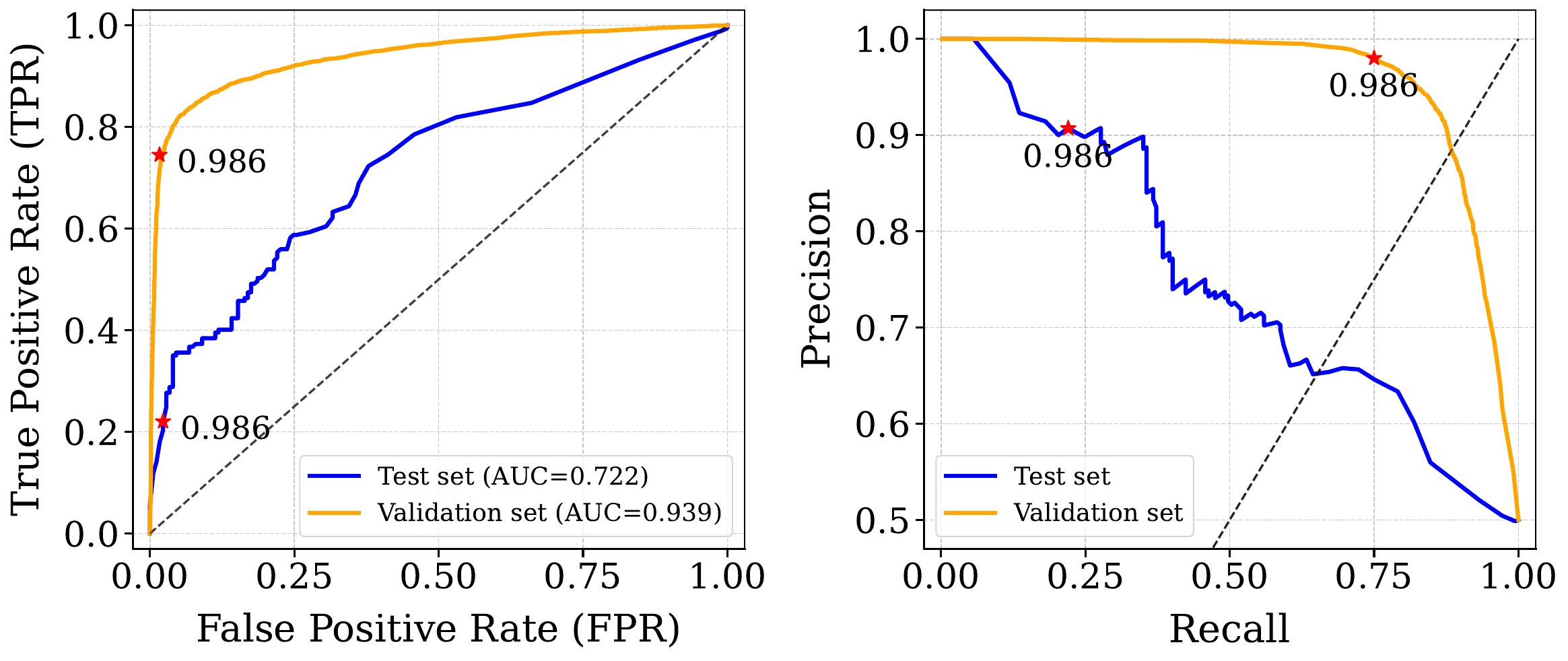}

        \caption{Comparison of the performance of our classifier evaluated with validation and testing sets separately. The left panel shows the ROC curve (orange line) of our classifier on the validation set containing our mock images with the area under the curve (AUC=0.939). And the ROC curve (blue line) of our classifier on the testing set containing the lens candidates from the COOL-LAMPS project \citep{2025ApJ...979..184M} with the area under the curve (AUC=0.722). The right panel shows the PR curve for the above two cases. We also mark the threshold (0.986) adopted in last iteration with red stars.}
        \label{fig:ROC}
    \end{figure}

Figure \ref{fig:loss} illustrates the variation of the loss function and AUC with its derivative across training epochs. In addition to the set of hyper-parameter configurations mentioned earlier  (Case 1), we also compared and used another set of hyper-parameters configuration (Case 2), which use an initial learning rate of $1\times10^{-4}$, a final learning rate of $1\times10^{-5}$, a weight decay of $1\times10^{-2}$ , with Cosine Annealing. It shows that in Case 1, after overcoming an initial local optimum, the AUC increases rapidly to approximately 0.93–0.95 before entering a period of fluctuation. However, the training loss begins to increase abnormally after 100 epochs, rendering the model unreliable beyond this point. In Case 2, the validation loss initially reached a relatively low level, but then gradually increased due to overfitting. Although the AUC in Case 2 reached a value of approximately 0.97, which was higher than that in Case 1, the highly fluctuating loss curve suggests that the model may be unstable. Moreover, the excessively high AUC also poses the risk of overfitting on the mock images. Therefore, we selected a deep learning model with an AUC of 0.94 from those in Case 1. The ROC curve of the best model on the validation set from one of our training sessions is shown in Figure \ref{fig:ROC}. On the simulated testing data, the model achieves an area under the curve (AUC) of 0.939.

We also define two additional metrics:
\begin{equation}
     \text{Precision} = \frac{\text{TP}}{\text{TP} + \text{FP}} , 
	\end{equation}
	\begin{equation}
     \text{Recall} = \text{TPR} = \frac{\text{TP}}{\text{TP} + \text{FN}} ,
\end{equation}
which are used to determine the probability threshold for human inspection. Figure \ref{fig:ROC} shows the Precision-Recall Curve (PRC) of the best model in one of our training sessions. In an ideal PRC, the curve would pass through the point (1, 1), where all thresholds would converge. The PRC is consulted to select the probability threshold for inspection, balancing precision (purity) and recall (completeness). For this work, we prioritized a high-precision point to demonstrate the model's accuracy and minimize the number of false positives, resulting 4,000 positive images.

To improve completeness while maintaining a low false-positive rate, we adopt an iterative search procedure for strong-lensing candidates. In the initial iteration, we manually filter the top 4,000 samples with the highest scores given by the classifier trained by the mock images described in Section \ref{subsec: mock}, and the corresponding threshold of the score is 0.99048. We then identified 102 true-positive candidates. After that, we removed them from the data pool for mining lens candidates and incorporated them into the positive sample of the training set for the subsequent training-filtering-updating loops. The details of the performance and outcomes of the classifier in each iteration are shown in Table \ref{tab:candidate_statistics}.

\begin{figure}[htb]

            \centering
            \includegraphics[width=1\linewidth]{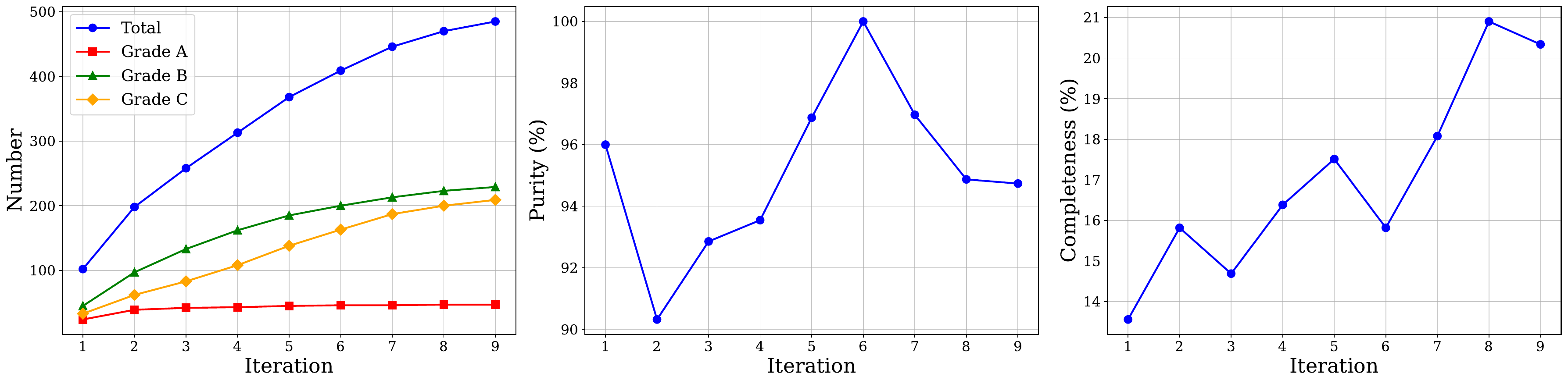}

        \caption{Performance of our classifier along with the interations. Left Panel shows the cumulative number at each iteration, shown for all candidates and separately for the three grades. The middle and right panels shows the Purity and completeness when applying our classifier to the testing set containing the lens candidates from the COOL-LAMPS project.}
        \label{fig:iteration}
    \end{figure}

Figure \ref{fig:iteration} shows the number of our candidates with grade A$\sim$C (see Section \ref{subsec:inf}). We repeat this iterative process until the number of detected strong-lensing candidates no longer shows significant improvement. Based on our experience, nine iterations are typically sufficient to achieve satisfactory results, since the quantity of the highest-quality grade A samples did not increase significantly in the final several iterations. This systematic refinement enriches the training data and enhances the model's robustness in identifying strong gravitational lenses. % other information

\begin{table}[htbp]
  \centering
  \caption{Statistics of candidates discovered during iterations (Iter.), including our threshold to select the top 4000 samples (Thresh.); the lowest score for our newly discovered candidates (Thresh.TP); the number of our newly discovered candidates (New) and cumulative candidates (Cum.), and the cumulative candidates with Grade A to Grade C; the purity (Purity) and completeness (Compl.) on the testing set made by COOL-LAMPS strong lensing \citep{2025ApJ...979..184M}.}

  \label{tab:candidate_statistics}
  \begin{tabular}{crrrrrrrrr}
    \hline
    \hline
    \textbf{Iter.} & \textbf{Thresh.} & \textbf{Thresh.TP} & \textbf{New} & \textbf{Cum.} & \textbf{Cum. A} & \textbf{Cum. B} & \textbf{Cum. C} & \textbf{Purity (\%)} & \textbf{Compl. (\%)} \\
    \hline
    1  & 0.99048 & 0.99062 & 102 & 102 & 24 & 45  & 33  & 96.0 & 13.6 \\
    \hline
    2  & 0.96557 & 0.96641 & 96 & 198 & 39 & 97  & 62  & 90.3 & 15.8 \\
    \hline
    3  & 0.97076 & 0.97081 & 60 & 258 & 42 & 133 & 83  & 92.9 & 14.7 \\
    \hline
    4  & 0.96627 & 0.96647 & 55 & 313 & 43 & 162 & 108 & 93.5 & 16.4 \\
    \hline
    5  & 0.97106 & 0.97146 & 55 & 368 & 45 & 185 & 138 & 96.9 & 17.5 \\
    \hline
    6  & 0.99082 & 0.99116 & 41 & 409 & 46 & 200 & 163 & 100.0 & 15.8 \\
    \hline
    7  & 0.97178 & 0.97300 & 37 & 446 & 46 & 213 & 187 & 97.0 & 18.1 \\
    \hline
    8  & 0.96549 & 0.96645 & 24 & 470 & 47 & 223 & 200 & 94.9 & 20.9 \\
    \hline
    9  & 0.98628 & 0.98748 & 15 & 485 & 47 & 229 & 209 & 94.7 & 20.3 \\
    \hline
    \hline
  \end{tabular}
\end{table}

\section{Results}\label{Sec: result}

In this section, we present our results. Subsection \ref{subsec:inf} introduces the grading system for the identified candidates and the distribution of their parameters. Subsection \ref{subsec:match} compares our findings with those of previous studies, and subsection \ref{subsec: Examples} highlights specific examples that demonstrate the effectiveness of our scoring system. Finally, subsection \ref{subsec: other} gives some other discussion.

\subsection{Inference and Lens Candidates}\label{subsec:inf}

Our analysis of 540,432 galaxy clusters from the \cite{Zou_2021} catalog identify a total of 485 strong gravitational lensing candidates using our trained classification model. These candidates are graded based on the prominence of their gravitational lensing features and galaxy cluster characteristics. The grading system is as follows:
\begin{enumerate}
    \item \textbf{Characteristics of galaxy clusters:} A system is considered to satisfy the requirements if the number of member galaxies in the field of view exceeds 10. Otherwise, it is regarded as lacking the distinct characteristics of a galaxy cluster.
    
    \item \textbf{Arc prominence:} The system must include a lensed arc that is clearly observable, with high relative brightness, a significant length-to-width ratio, and a high degree of circumferential completeness. Arcs that fail to meet all these criteria are considered non-prominent.
    
\end{enumerate}
The scoring rules are defined as follows:
\begin{enumerate}
    \item If Criterion 1 is satisfied:
    \begin{enumerate}
        \item and if Criterion 2 is also satisfied, the score is \textbf{A}.
        \item and if Criterion 2 is not satisfied but an arc is visual, the score is \textbf{B}.
        \item and if Criterion 2 is not satisfied and the arc is nearly invisible, the score is \textbf{C}.
    \end{enumerate}
    \item If Criterion 1 is not satisfied: \begin{enumerate}
        \item and if Criterion 2 is satisfied, the score is \textbf{B}.
        \item and if Criterion 2 is not satisfied, the score is \textbf{C}.
    \end{enumerate}
\end{enumerate}

\begin{figure}[htb]
        \centering
        \includegraphics[scale=0.4]{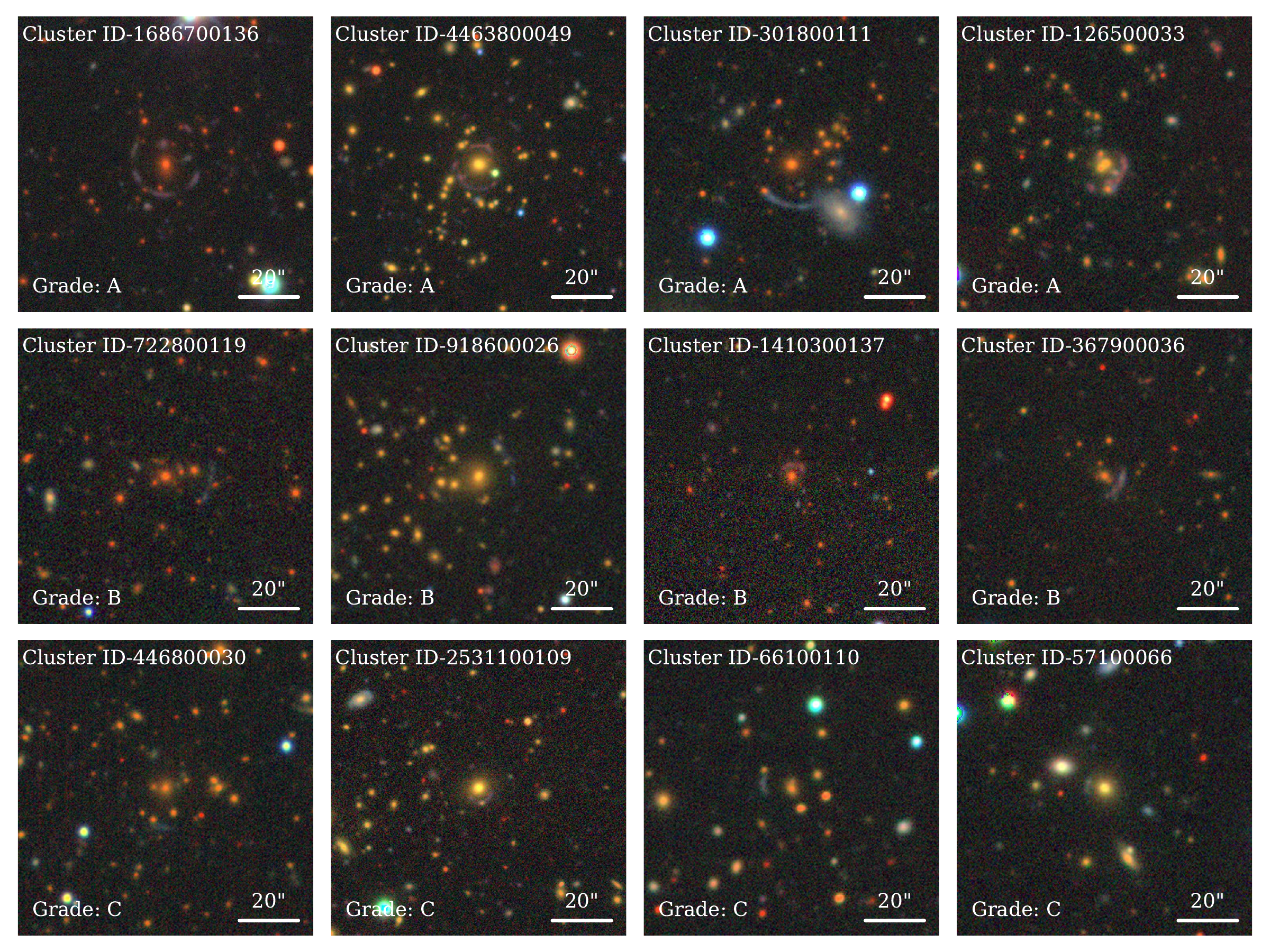}

        \caption{Selected twelve typical strong lensing candidates with Grade A, B and C. Grade A systems exhibit prominent, long, and bright arc-like images, with distinct features of background galaxy clusters. In contrast, the arcs in Grade B and Grade C systems are fainter. For all images, north is up, and east to the left. The top left corner of each image is the galaxy cluster ID of each image in \citealt{Zou_2021}; the grade is indicated at the bottom left corner.}

        \label{fig:example}
    \end{figure}

\begin{figure}[htb]
        \centering
        \includegraphics[scale=0.65]{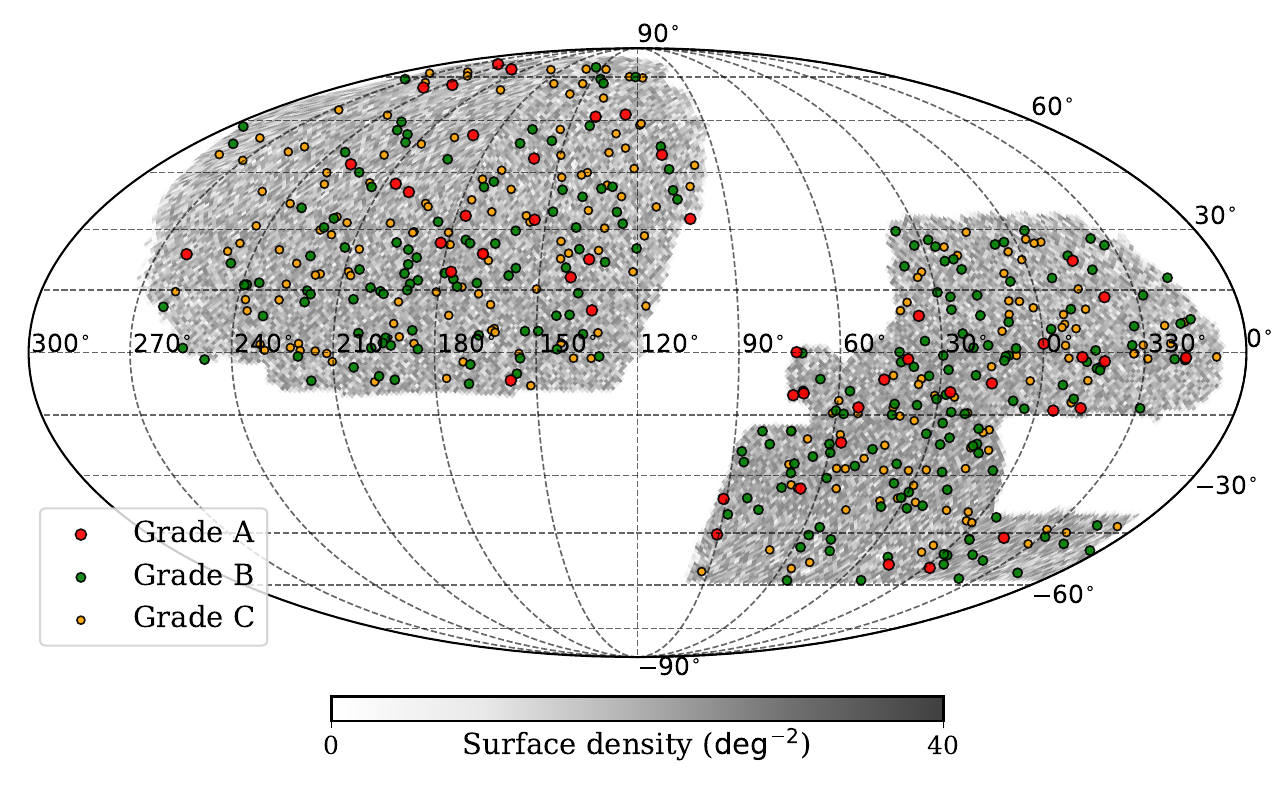}
        
        \caption{The 485 candidates over the surface density map of clusters in the catalog from \citealt{Zou_2021}.}
        \label{fig:result_pos}
    \end{figure}

After all seven authors have completed their respective grading, we canceling out A and C grades against each other. With seven scoring experts, the remaining number of grades is always odd and consists only of A and B or B and C. In this case, the majority grade is selected as the final grade (for example, 4A, 2B, 1C; after cancellation, it becomes 3A, 2B, and A is chosen). We finally identify 47 Grade A candidates (9.7\%), 229 Grade B candidates (47.2\%), and 209 Grade C (43.1\%) candidates. Examples of candidates from each grade are shown in Figure \ref{fig:example}, while the positions of all candidates in the sky are displayed in Figure \ref{fig:result_pos}. 

\begin{figure}[htb]
        \centering
        \includegraphics[scale=0.45]{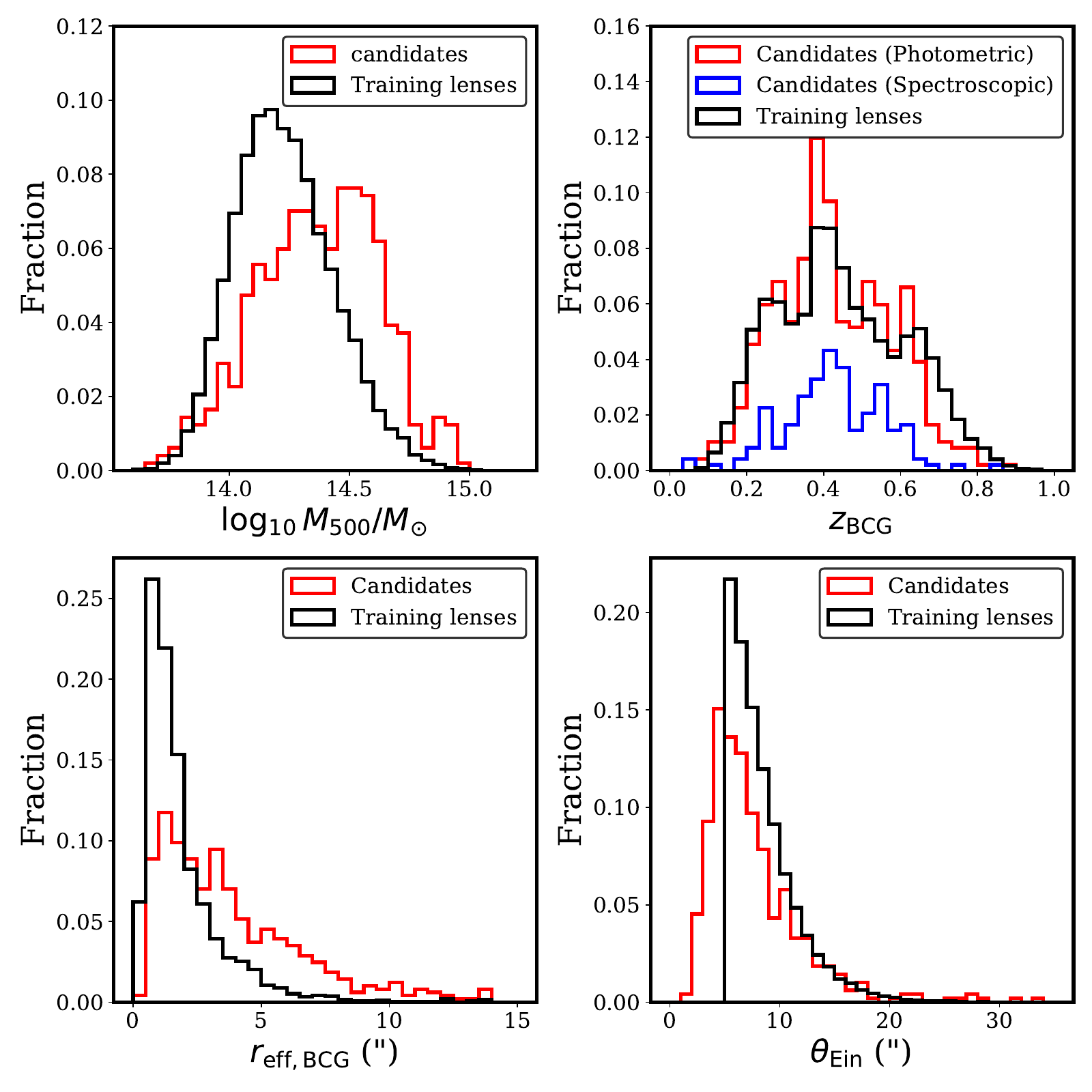}
        
        \caption{Parameters distribution of our strong lensing candidates and training samlpes. Upper left: The $M_{500}$ of the cluster. Upper right: The redshift of BCGs, including the photometric redshifts of all candidates (red), as well as the subset of candidates whose spectroscopic redshifts can be found in DESI DR1 (\citealt{desicollaboration2025datarelease1dark}, in purple). Lower left: The effective radius of BCG. Lower right: The Einstein radius, which is approximately estimated by taking the average of the distances from the BCG to several points on the arcs.}
        \label{fig:dis}
    \end{figure}

\begin{figure}[htb]
        \centering
        \includegraphics[scale=0.4]{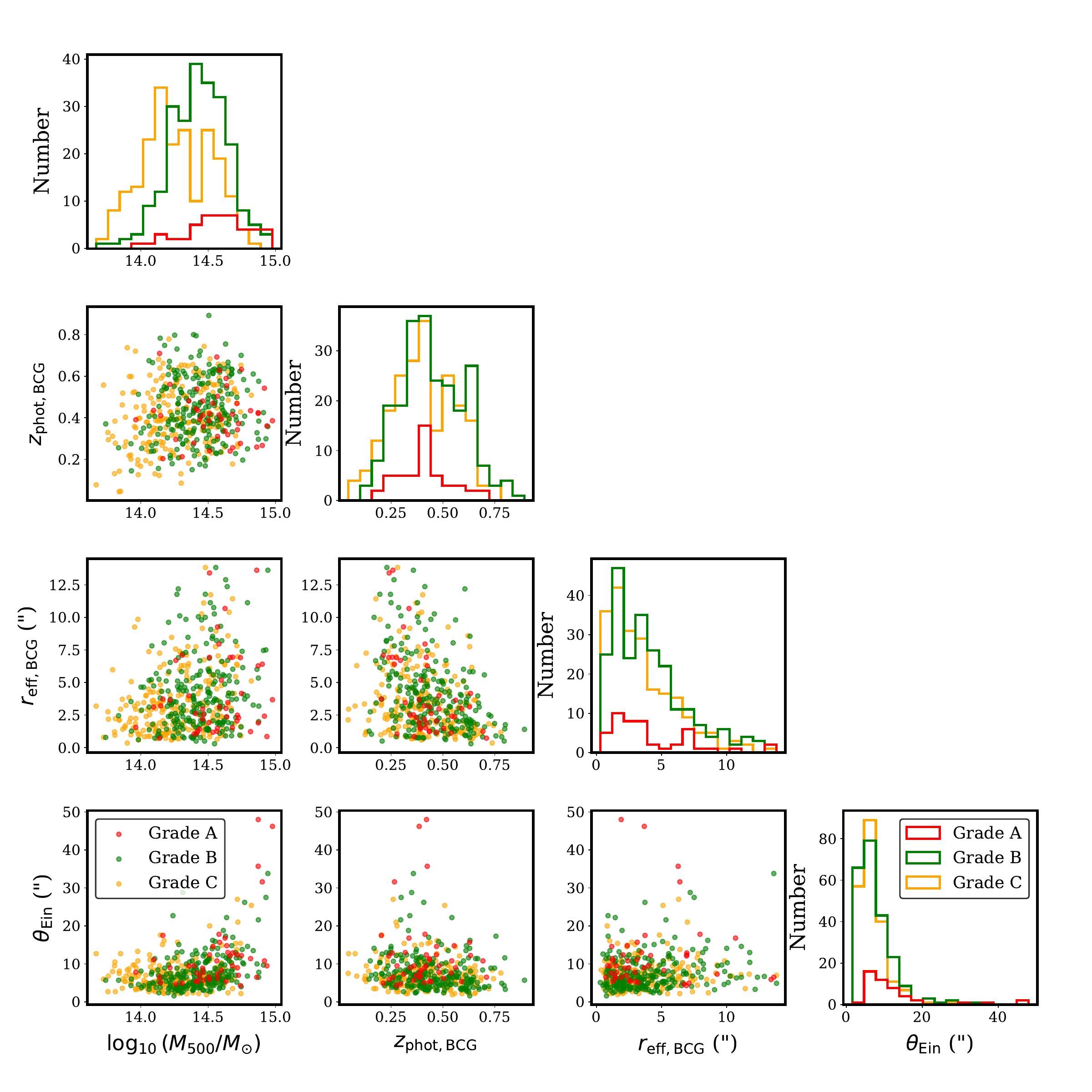}

        \caption{Corner plots illustrate the distributions and correlations among four key parameters for three graded strong lensing candidates displaying scatter plots with Grade A (red), Grade B (green), and Grade C (orange) clusters. The diagonal presents the parameter distributions. The four parameters are: (a) cluster mass $M_{500}$, (b) Photometric redshift of BCG, (c) Effective of BCG, and (d) Einstein Radius. }
        \label{fig:cornerPlots}
    \end{figure}

Our systematic analysis of lens candidate properties reveals key demographic trends (Figure \ref{fig:dis}). The cluster masses $M_{500}$, derived from \cite{Zou_2021} catalog through X-ray and radio observations, exhibit a distribution range of $10^{13.67\sim 14.97}M_{\odot}$, which is consistent with our training samples. However, the peak mass of our candidates ($\sim10^{14.6}M_{\odot}$) is higher than that of the training set ($\sim10^{14.2}M_{\odot}$). The figure also illustrates the distribution of the BCG photometric redshift and effective radius of our candidates, which align closely with the distribution in our training samples. Notably, the Einstein ring radius distribution differs. While the training set includes only samples with Einstein radii greater than $5''$ our candidates include some with smaller Einstein rings. It is important to note that the Einstein ring radii here are measured approximately as the distance from the arc to the lens center, which differs from the definition used in the training set. Furthermore, although our model is designed to identify strong lensing with prominent arcs at galaxy cluster scales, it is expected to detect some by-products at galaxy or group scales. In Figure \ref{fig:cornerPlots}, we show a corner plot for the correlations among these four parameters, and also labeled the grade of these candidates assigned by authors, showing that candidates with Grade A have higher $M_{500}$ than others.

\subsection{Comparison with other strong lensing catalogs}\label{subsec:match}

We compared our results with other studies that identified strong lensing candidates in DESI data. The first comparison is with a series of paper by \cite{Huang_2020}, \cite{2021ApJ...909...27H}, and \cite{2024ApJS..274...16S}, which utilized deep residual neural network to search for strong lensing in DESI Legacy Imaging Surveys DR8 and DR9. The second comparison is with the project ChicagO Optically selected Lenses Located at the Margins of Public Surveys (COOL-LAMPS; \citealt{2025ApJ...979..184M}), which relied on a large team of researchers to manually perform visual searches for strong gravitational lenses in DESI data.

%Cumulatively, 172 candidates remain undetected by both surveys, comprising 12 Grade A (4.9\%), 96 Grade B (38.8\%), and 139 Grade C (56.3\%) systems.

In the recent work by \cite{2024ApJS..274...16S} of the first series work, they found a total of 3,057 strong lens candidates with Grades A, B and C, and also Grade D candidates which they did not count in their paper but included in their project website. From their catalog, we recover 227 previously known systems and identify 258 new candidates. These new detections include 20 candidates of Grade A (7.8\%), 96 candidates of Grade B (37.2\%) and 142 candidates of Grade C (55.0\%). 

\begin{figure}[htb]
        \centering
        \includegraphics[scale=0.45]{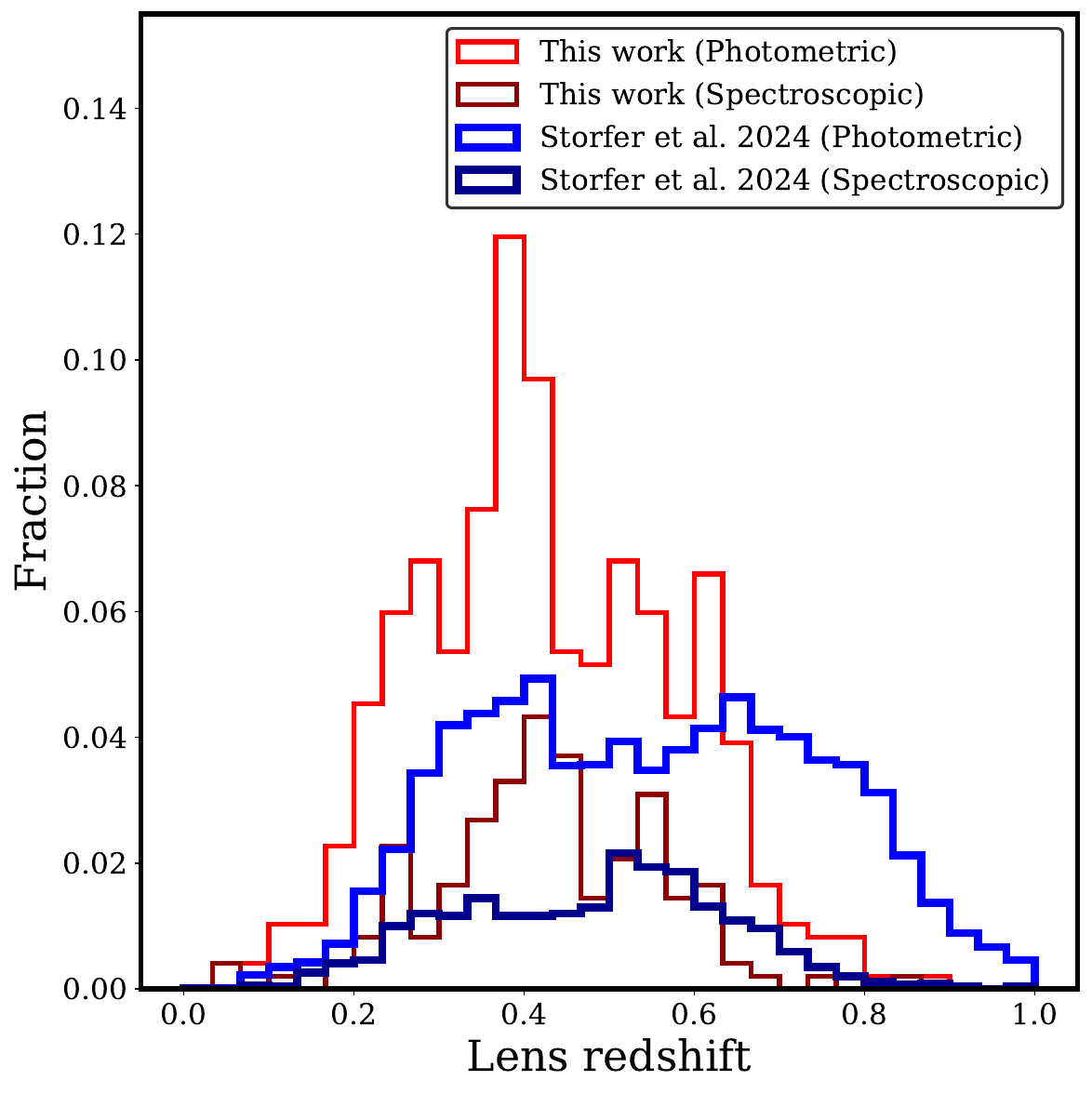}

        \caption{Lens redshift of candidates from our work and \citealt{2024ApJS..274...16S}. For our work, we show the photometric redshift for all the candidates (red), as well as the subset of candidates whose spectroscopic redshift can be found in DESI DR1 (purple). For \citealt{2024ApJS..274...16S}, we show either photometric (black line) or spectroscopic from SDSS Data Release 17 (blue line).}
        \label{fig:huang z}
    \end{figure}

Figure \ref{fig:huang z} demonstrates that our candidates exhibit a higher number density at low photometric redshifts. This discrepancy could be attributed to the smaller Einstein ring radii typically associated with low-redshift galaxy-scale lenses, which are more challenging to detect. In contrast, our CNN architecture, optimized specifically for cluster-scale lenses with higher Einstein radii, is less affected by this limitation. An analysis of some candidates from \cite{2024ApJS..274...16S} reveals masses within the Einstein radius of approximately $\sim 2\times 10^{12} M_{\odot}$,  characteristic of galaxy-scale systems. This highlights our method's enhanced sensitivity to identifying lensing systems at the galaxy cluster scale.

For the overlapping candidates, we assigned 27 to Grade A (11.9\%), 133 to Grade B (58.6\%) and 67 to Grade C (29.5\%), while \cite{2024ApJS..274...16S} gave 106 Grade A (46.7\%), 47 Grade B (20.7\%), 52 Grade C (22.9\%) and 22 Grade D (9.7\%). Our grading strategy not only accounts for the strength of the gravitational lensing signal but also considers the significance of the galaxy cluster characteristics. As a result, our grades are generally lower than those given by \cite{2024ApJS..274...16S}. However, there are exceptional cases where our grades are higher. We will briefly discuss these cases later in Section \ref{subsec: Examples}.

\begin{figure}[htb]
        \centering
        \includegraphics[scale=0.4]{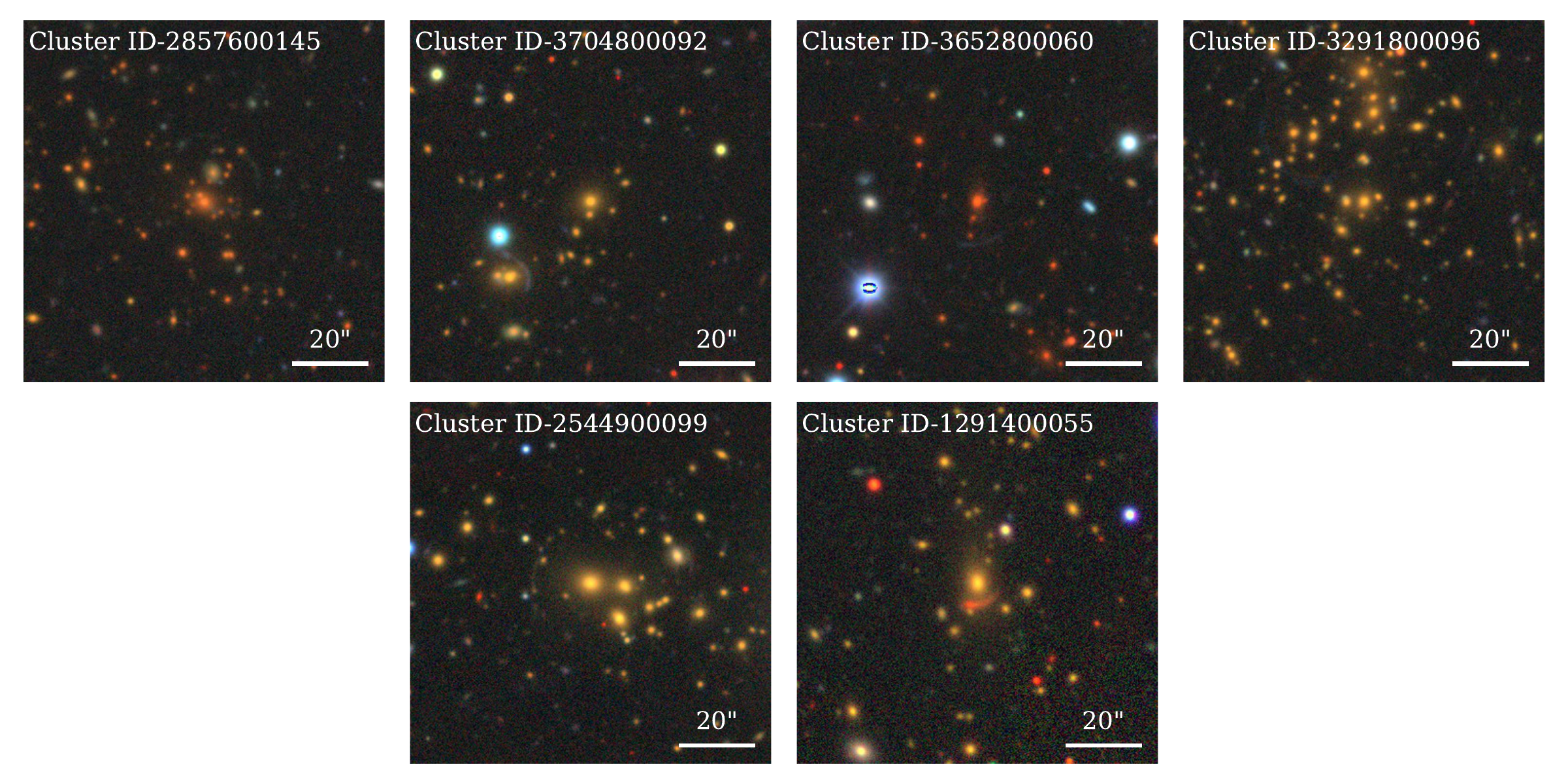}

        \caption{Six strong lensing systems identified in project COOL-LAMPS but were not find by this work. The arcs in this six images are generally fainter compared to our Grade A candidates, and the centers of the arcs in the upper-middle and lower-left images do not point towards the BCG of the cluster. Other systems with even low quality are not shown here.}
        \label{fig:cool-lamps}
    \end{figure}

We also compare our findings with another project called COOL-LAMPS finding cluster-scale strong gravitational lenses. In their recent paper \citep{2025ApJ...979..184M}, they report a total of 177 cluster-scale strong gravitational lens systems, and all of these systems are included in \cite{Zou_2021}. Of our 485 systems, 429 are newly identified by our work, comprising 32 Grade A (7.5\%), 193 Grade B (45.0\%) and 204 Grade C (47.5\%). For the overlapping candidates, we have 15 Grade A (26.8\%), 36 Grade B (64.3\%) and 5 Grade C (8.9\%), indicating that our machine learning based method can complement the manual searches. We also examined 122 COOL-LAMPS candidates that our method failed to detect. The vast majority of these images would receive a low grade according to our grading strategy. Only fewer than 6 samples (Figure \ref{fig:cool-lamps}) could receive a Grade A, demonstrating that our method has good completeness in identifying candidates at the Grade A level.

Based on the matching of the two projects, there are a total of 247 candidates among our candidates who are not discovered by them. Among them, 16 are Grade A (6.5\%), 90 are Grade B (36.4\%), and 141 are Grade C (57.1\%).

\subsection{Grading strategy}\label{subsec: Examples}

% candidates that we have a discrete Grade.

\begin{figure}[htb]
        \centering
        \includegraphics[scale=0.4]{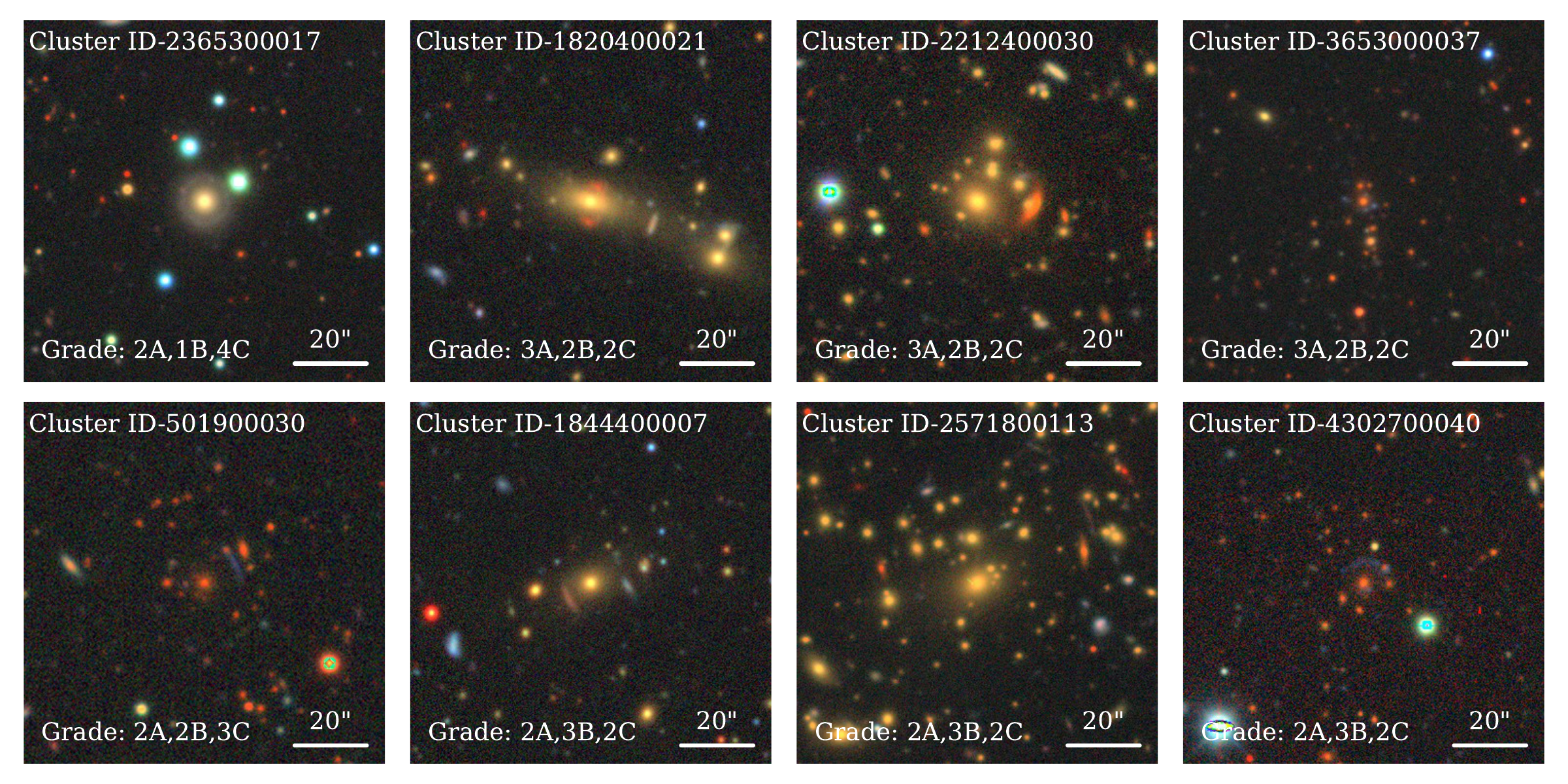}

        \caption{Eight strong lensing candidates with the largest standard deviation among the grades by 7 authors (we give A a score of 3, B of 2, and C of 1). For all images, north is up, and east to the left. The top left corner of each image is the galaxy cluster ID of each image in \citealt{Zou_2021}; bottom left corner, the grades number by 7 authors. }
        \label{fig:exp_scatter}
    \end{figure}

We present a few examples with our grading strategy whose grades show significant variation among the authors. In Figure \ref{fig:exp_scatter} we highlight the eight candidates with the largest standard deviations. The galaxy cluster 2365300017 (first row, first column) has a ring-galaxy like BCG, which also looks like an Einstein ring. The color of the ring appears quite similar to that of BCG, and such a perfect Einstein ring is very rare among strong-lensing systems. We believe the final Grade C, assigned based on our third grading strategy, is more reliable than the Grade B given by the other two strategies  (see Section \ref{subsec:inf}). For the other seven images, the arcs are noticeably different in color from the member galaxies, making them appear observable. This led some of the authors to give Grade A. However, compared to other strong lensing candidates, these arcs are relatively short and faint. As a result, all of them were ultimately assigned Grade B as a compromise.

In Section \ref{subsec:match}, we note several cases where the grades in this work are higher than those in \cite{2024ApJS..274...16S}, which is unusual. For instance, the galaxy cluster 2498800056, also labeled Abell 370 in \cite{1958ApJS....3..211A}, features a giant arc which is about $40''$ far from its BCG, and is Grade A in this work and Grade D in \cite{2024ApJS..274...16S}. In the training set and recognition work of our network, we have cropped the image side length to $60''$, which cannot catch that giant arc. However, the images used in our grading work are not cropped, allowing all authors assigned Grade A to this candidate. The image cutout size in \cite{2024ApJS..274...16S} is even smaller than us ($26''$), which could not capture obvious strong lensing signal, and may be the reason why they gave a low grade. Machine learning methods can capture some gravitational lensing features that are difficult to identify visually This might be the reason why it can identify these candidates even when the image dimensions are relatively small.

% other special candidates.

\subsection{Other discussion}\label{subsec: other}

In the process of searching, the vast majority of false positives consist of the images with similar features to the strong lensed arcs for which the specific cause of misclassification remains unclear, suggesting the deep learning model may be sensitive to subtle or complex morphological patterns. However, during the iteractive procedure, we can identify and categorize a subset of significant false positives attributable to recognizable features, including the spiral arms of late-type galaxies, edge-on disk galaxies, ringed galaxies, and stellar streams, all of which can produce arc-like features. Additionally, the background galaxies with significant photometric color contrast against the central brightest cluster galaxy (BCG) can occasionally generate false lensing signals of multiple lensed images, which can not be determined with images only but request further spectrum information. % Representative examples of these identifiable systematic false positives are presented in Figure \ref{}.

It is worth noting that modern architectures such as EfficientNet \citep{tan2020efficientnetrethinkingmodelscaling}, ConvNeXt \citep{liu2022convnet2020s}, and Vision Transformers \citep{dosovitskiy2021imageworth16x16words}. are not used in this work, since the core objective of this study is not to compare state-of-the-art networks, but to systematically examine whether human-in-the-loop collaboration can enhance detection performance even with a well-established baseline model. We selected a modified ResNet-18 for its stability, interpretability, and wide adoption in earlier lens-finding studies (\citealt{2018MNRAS.473.3895L, 2021ApJ...923...16L, 2019MNRAS.484.3879P}), which provides a consistent and reproducible foundation for evaluating the added value of human expertise. While we fully acknowledge the strong potential of newer architectures (\citealt{2025MNRAS.tmp.1632P,2025A&A...702A.130N,2025PASP..137f4504Y}) and agree they represent a promising future direction. Also, effectively integrating them with human-in-the-loop interaction requires careful design to ensure meaningful synergy between model capacity and human judgment. Nevertheless, we plan to explore its extension to more advanced architectures in subsequent research, where both model selection and human integration can be co-optimized.

 \section{Summary}\label{Sec: discussion}

In this work, we present a systematic search for candidates of galaxy cluster-scale strong gravitational lensing systems based on the galaxy cluster catalog identified by \cite{Zou_2021} from the DESI Legacy Imaging Surveys DR8, covering approximately $20,000 \  \rm deg^{2}$ of the extragalactic sky. We first designed a 3-step approach, including a strong lens simulator for creating positive samples in the initial training set, a binary classifier built upon Resnet-18 for identifying lens candidates, and human inspection to improve the purity of the outcomes further. Then, we interactively loop the classification and human inspection procedures to achieve stable completeness of the lens candidates. Subsequently, all the authors of this work visually assessed the candidates according to the features of strong lensing and galaxy clusters. As a result, we identified 485 high-confidence cluster lens candidates. The halo mass, $\log(M_{500}/M_{\odot})$, of these candidates ranges from $10^{13.67}$ to $10^{14.97} M_{\odot}$, and their brightest cluster galaxy (BCG) photometric redshift spans from 0.04 to 0.89.

The iteration starts with applying a binary classifier trained with the initial training sets (50,000 P + 50,000 N) to galaxy clusters from DESI Legacy Survey DR8. We chose a relatively conservative configuration of hyperparameters (a batch size of 32, a learning rate of $3\times10^{-6}$, and a weight decay of $1\times10^{-1}$). The training process continues until the point where the Area Under the ROC Curve (AUC) reaches its maximum value, after which the model's loss begins to increase abnormally, approximately 0.94. Although the AUC of the model we selected was not as high as that of the more aggressive configuration (a batch size of 32, an adaptive learning rate with an initial value of $1\times10^{-4}$ and final value of $1\times10^{-5}$, and a weight decay of $1\times10^{-2}$, with Cosine Annealing), leading to an AUC of 0.97, more robust evaluation beyond AUC is required to avoid the potential overfitting issue and find the optimal configuration of the hyperparameters for the entire process, which is one of our focuses in the furture studies. After that, the first author visually selected candidates from the 4,000 samples with the highest scores given by the model, and added them to the original training set. The classifier is then retrained using this updated dataset. The loop will stop when the classification performance is stable (see Figure \ref{fig:iteration}). We implement nine iterations in this study. At last, through visual inspection using a three-tier grading system based on arc morphology and Einstein radius scale (criteria detailed in Section \ref{Sec: result}) by all the authors of this paper, we classify 47 candidates as grade A (obvious arcs in cluster scale), 229 as grade B (probable arcs or galaxy scale), and 209 as grade C (possible arcs and galaxy scale).

In comparison to relevant studies, our strategy emphasizes the identification of cluster lens candidates in a Human-in-the-loop-like manner \citep{2024ChPhC..48i5001F}. First, relative to \cite{2024ApJS..274...16S} 's CNN-based galaxy-scale lens search, our training set is optimized based on characteristics of cluster lenses (see Subsection \ref{subsec: mock}), which is a key reason why we found hundreds of novel cluster lens candidates. Second, compared with the human-intensive method of COOL-LAMPS \citep{2025ApJ...979..184M}, our CNN-based approach achieves a computational efficiency of $ < \rm 10 ms$ per image classification, demonstrating the viability of machine learning for large-area surveys. Furthermore, our strategy achieves a simple human-computer interaction by retraining the model through iterative processes, incorporating manually identified true positive samples into the training set. This approach balances the accuracy of manual search and the speed of machine learning methods. Cross-matching with existing catalogs reveals that 247 candidates (50.9\% of the sample) show no counterparts in \cite{2024ApJS..274...16S} or the COOL-LAMPS project \citep{2025ApJ...979..184M}, which include 16 new grade A systems and 90 grade B, 141 grade C.

However, several aspects of our approach require improvement. First, there are discrepancies between the distributions of the training sets and those of the candidates (see Figure \ref{fig:dis}), which suggests that we should incorporate more realistic parametric distributions when generating training sets to enhance the classification performance of the deep learning module. Second, the current ResNet-18 achieves an AUC of $\gtrsim0.94$ in validation tests at each step, while advanced architectures like Vision Transformers \citep{dosovitskiy2021imageworth16x16words} have the potential to improve the AUC to 0.999 through better arc morphology capture. Thus, it is sensible to explore novel AI techniques in future investigations. Third, our manual verification process (about 2.5 hours per 1000 candidates) in the iterative process could be streamlined using active learning techniques, such as Bayesian active learning \citep{2011arXiv1112.5745H}, to prioritize ambiguous cases, potentially cutting human effort by 50\%.

To summarize, with a 3-step lens-finding approach, we discover 485 high-confidence cluster lens candidates in total cluster lens candidates from DESI Legacy Surveys DR8, and 247 of them are new. This dataset is valuable for examining mass assembly histories through joint strong and weak lensing analyses. Among the findings, 16 Grade A systems are suitable for putting strong constraints on the concentration-mass relation of dark matter halos and the cosmological parameters, such as the matter density, $\Omega_m$. In addition, comparing observations taken at different epochs could be beneficial in detecting lensed transients. We encourage observers to explore the best candidates for these applications in the future.

\section*{Acknowledgements}

Z.Z. acknowledges the High-performance Computing center at Westlake University and the Tsinghua Astrophysics High-Performance Computing platform for providing computational and data storage resources that have contributed to the research results reported in this work. N.L. acknowledges the support of the CAS Project for Young Scientists in Basic Research (No. YSBR-062), the Ministry of Science and Technology of China (No. 2020SKA0110100), and the support of the Association for Astronomy X A.I. $(A^3)$, funded by the Science and Education Integration Funding of the University of Chinese Academy of Sciences. Z.H. acknowledges support from the National Natural Science Foundation of China (Grant No. 12403104). H.Z. acknowledges the supports from the National Key R\&D Program of China (grant Nos. 2023YFA1607804 and 2022YFA1602902) and the National Natural Science Foundation of China (NSFC; grant Nos. 12120101003 and 12373010).

\bibliography{ref}{}

@article{Zou_2021,
   title={Galaxy Clusters from the DESI Legacy Imaging Surveys. I. Cluster Detection},
   volume={253},
   ISSN={1538-4365},
   url={http://dx.doi.org/10.3847/1538-4365/abe5b0},
   DOI={10.3847/1538-4365/abe5b0},
   number={2},
   journal={The Astrophysical Journal Supplement Series},
   publisher={American Astronomical Society},
   author={Zou, Hu and Gao, Jinghua and Xu, Xin and Zhou, Xu and Ma, Jun and Zhou, Zhimin and Zhang, Tianmeng and Nie, Jundan and Wang, Jiali and Xue, Suijian},
   year={2021},
   month=apr, pages={56} }

@article{Zou_2019,
doi = {10.3847/1538-4365/ab1847},
url = {https://dx.doi.org/10.3847/1538-4365/ab1847},
year = {2019},
month = {may},
publisher = {The American Astronomical Society},
volume = {242},
number = {1},
pages = {8},
author = {Hu Zou and Jinghua Gao and Xu Zhou and Xu Kong},
title = {Photometric Redshifts and Stellar Masses for Galaxies from the DESI Legacy Imaging Surveys},
journal = {The Astrophysical Journal Supplement Series}
}

@ARTICLE{2015ApJ...811...20C,
       author = {{Collett}, Thomas E.},
        title = "{The Population of Galaxy-Galaxy Strong Lenses in Forthcoming Optical Imaging Surveys}",
      journal = {\apj},
         year = 2015,
        month = sep,
       volume = {811},
       number = {1},
          eid = {20},
        pages = {20},
          doi = {10.1088/0004-637X/811/1/20},
archivePrefix = {arXiv},
       eprint = {1507.02657},
 primaryClass = {astro-ph.CO},
       adsurl = {https://ui.adsabs.harvard.edu/abs/2015ApJ...811...20C}
}

@article{Huang_2020,
   title={Finding Strong Gravitational Lenses in the DESI DECam Legacy Survey},
   volume={894},
   ISSN={1538-4357},
   url={http://dx.doi.org/10.3847/1538-4357/ab7ffb},
   DOI={10.3847/1538-4357/ab7ffb},
   number={1},
   journal={The Astrophysical Journal},
   publisher={American Astronomical Society},
   author={Huang, X. and Storfer, C. and Ravi, V. and Pilon, A. and Domingo, M. and Schlegel, D. J. and Bailey, S. and Dey, A. and Gupta, R. R. and Herrera, D. and Juneau, S. and Landriau, M. and Lang, D. and Meisner, A. and Moustakas, J. and Myers, A. D. and Schlafly, E. F. and Valdes, F. and Weaver, B. A. and Yang, J. and Yèche, C.},
   year={2020},
   month=may, pages={78} }

@misc{he2015deepresiduallearningimage,
      title={Deep Residual Learning for Image Recognition}, 
      author={Kaiming He and Xiangyu Zhang and Shaoqing Ren and Jian Sun},
      year={2015},
      eprint={1512.03385},
      archivePrefix={arXiv},
      primaryClass={cs.CV},
      url={https://arxiv.org/abs/1512.03385}, 
}

@ARTICLE{2017PASP..129f4101Z,
       author = {{Zou}, Hu and {Zhou}, Xu and {Fan}, Xiaohui and {Zhang}, Tianmeng and {Zhou}, Zhimin and {Nie}, Jundan and {Peng}, Xiyan and {McGreer}, Ian and {Jiang}, Linhua and {Dey}, Arjun and {Fan}, Dongwei and {He}, Boliang and {Jiang}, Zhaoji and {Lang}, Dustin and {Lesser}, Michael and {Ma}, Jun and {Mao}, Shude and {Schlegel}, David and {Wang}, Jiali},
        title = "{Project Overview of the Beijing-Arizona Sky Survey}",
      journal = {\pasp},
         year = 2017,
        month = jun,
       volume = {129},
       number = {976},
        pages = {064101},
          doi = {10.1088/1538-3873/aa65ba},
archivePrefix = {arXiv},
       eprint = {1702.03653},
 primaryClass = {astro-ph.GA},
       adsurl = {https://ui.adsabs.harvard.edu/abs/2017PASP..129f4101Z}
}

@INPROCEEDINGS{2016AAS...22831701B,
       author = {{Blum}, Robert D. and {Burleigh}, Kaylan and {Dey}, Arjun and {Schlegel}, David J. and {Meisner}, Aaron M. and {Levi}, Michael and {Myers}, Adam D. and {Lang}, Dustin and {Moustakas}, John and {Patej}, Anna and {Valdes}, Francisco and {Kneib}, Jean-Paul and {Huanyuan}, Shan and {Nord}, Brian and {Olsen}, Knut A. and {Delubac}, Timoth{\'e}e and {Saha}, Abi and {James}, David and {Walker}, Alistair R. and {DECaLS Team}},
        title = "{The DECam Legacy Survey}",
    booktitle = {American Astronomical Society Meeting Abstracts \#228},
         year = 2016,
       series = {American Astronomical Society Meeting Abstracts},
       volume = {228},
        month = jun,
          eid = {317.01},
        pages = {317.01},
       adsurl = {https://ui.adsabs.harvard.edu/abs/2016AAS...22831701B}
}

@INPROCEEDINGS{2016AAS...22831702S,
       author = {{Silva}, David R. and {Blum}, Robert D. and {Allen}, Lori and {Dey}, Arjun and {Schlegel}, David J. and {Lang}, Dustin and {Moustakas}, John and {Meisner}, Aaron M. and {Valdes}, Francisco and {Patej}, Anna and {Myers}, Adam D. and {Sprayberry}, David and {Saha}, Abi and {Olsen}, Knut A. and {Gaines}, Sasha and {Yang}, Qian and {Burleigh}, Kaylan J. and {MzLS Team}},
        title = "{The Mayall z-band Legacy Survey}",
    booktitle = {American Astronomical Society Meeting Abstracts \#228},
         year = 2016,
       series = {American Astronomical Society Meeting Abstracts},
       volume = {228},
        month = jun,
          eid = {317.02},
        pages = {317.02},
       adsurl = {https://ui.adsabs.harvard.edu/abs/2016AAS...22831702S}
}

@ARTICLE{2014Sci...344.1492R,
       author = {{Rodriguez}, Alex and {Laio}, Alessandro},
        title = "{Clustering by fast search and find of density peaks}",
      journal = {Science},
         year = 2014,
        month = jun,
       volume = {344},
       number = {6191},
        pages = {1492-1496},
          doi = {10.1126/science.1242072},
       adsurl = {https://ui.adsabs.harvard.edu/abs/2014Sci...344.1492R}
}

@ARTICLE{2020PASP..132b4101G,
       author = {{Gao}, Jinghua and {Zou}, Hu and {Zhou}, Xu and {Kong}, Xu},
        title = "{A Catalog of Galaxy Clusters Identified from SCUSS, SDSS, and UNWISE}",
      journal = {\pasp},
         year = 2020,
        month = feb,
       volume = {132},
       number = {1008},
          eid = {024101},
        pages = {024101},
          doi = {10.1088/1538-3873/ab6151},
archivePrefix = {arXiv},
       eprint = {1912.10909},
 primaryClass = {astro-ph.GA},
       adsurl = {https://ui.adsabs.harvard.edu/abs/2020PASP..132b4101G}
}

@ARTICLE{1996ApJ...462..563N,
       author = {{Navarro}, Julio F. and {Frenk}, Carlos S. and {White}, Simon D.~M.},
        title = "{The Structure of Cold Dark Matter Halos}",
      journal = {\apj},
         year = 1996,
        month = may,
       volume = {462},
        pages = {563},
          doi = {10.1086/177173},
archivePrefix = {arXiv},
       eprint = {astro-ph/9508025},
 primaryClass = {astro-ph},
       adsurl = {https://ui.adsabs.harvard.edu/abs/1996ApJ...462..563N}
}

@ARTICLE{2024ApJS..274...16S,
       author = {{Storfer}, C. and {Huang}, X. and {Gu}, A. and {Sheu}, W. and {Banka}, S. and {Dey}, A. and {Inchausti Reyes}, J. and {Jain}, A. and {Kwon}, K.~J. and {Lang}, D. and {Lee}, V. and {Meisner}, A. and {Moustakas}, J. and {Myers}, A.~D. and {Tabares-Tarquinio}, S. and {Schlafly}, E.~F. and {Schlegel}, D.~J.},
        title = "{New Strong Gravitational Lenses from the DESI Legacy Imaging Surveys Data Release 9}",
      journal = {\apjs},
         year = 2024,
        month = sep,
       volume = {274},
       number = {1},
          eid = {16},
        pages = {16},
          doi = {10.3847/1538-4365/ad527e},
archivePrefix = {arXiv},
       eprint = {2206.02764},
 primaryClass = {astro-ph.CO},
       adsurl = {https://ui.adsabs.harvard.edu/abs/2024ApJS..274...16S},
    shorthand = {S24}
}

@ARTICLE{2025ApJ...979..184M,
       author = {{Mork}, Simon D. and {Gladders}, Michael D. and {Khullar}, Gourav and {Sharon}, Keren and {Chicoine}, Nathalie and {Cloonan}, Aidan P. and {Dahle}, H{\r{a}}kon and {Garza}, Diego and {Glusman}, Rowen and {Gozman}, Katya and {Horwath}, Gabriela and {Levine}, Benjamin C. and {Liang}, Olina and {Mahronic}, Daniel and {Manwadkar}, Viraj and {Martinez}, Michael N. and {Masegian}, Alexandra and {Matthews Acu{\~n}a}, Owen S. and {Merz}, Kaiya and {Pan}, Yue and {Sanchez}, Jorge A. and {Sierra}, Isaac and {Kavin Stein}, Daniel J. and {Sukay}, Ezra and {Tamargo-Arizmendi}, Marcos and {Tavangar}, Kiyan and {Tu}, Ruoyang and {Wagner}, Grace and {Zaborowski}, Erik A. and {Zhang}, Yunchong and {Cool-Lamps Collaboration}},
        title = "{COOL-LAMPS. VII. Quantifying Strong-lens Scaling Relations with 177 Cluster-scale Strong Gravitational Lenses in DECaLS}",
      journal = {\apj},
         year = 2025,
        month = feb,
       volume = {979},
       number = {2},
          eid = {184},
        pages = {184},
          doi = {10.3847/1538-4357/ada24c},
archivePrefix = {arXiv},
       eprint = {2401.08575},
 primaryClass = {astro-ph.GA},
       adsurl = {https://ui.adsabs.harvard.edu/abs/2025ApJ...979..184M}
}

@ARTICLE{2016AJ....151...36L,
       author = {{Lang}, Dustin and {Hogg}, David W. and {Schlegel}, David J.},
        title = "{WISE Photometry for 400 Million SDSS Sources}",
      journal = {\aj},
         year = 2016,
        month = feb,
       volume = {151},
       number = {2},
          eid = {36},
        pages = {36},
          doi = {10.3847/0004-6256/151/2/36},
archivePrefix = {arXiv},
       eprint = {1410.7397},
 primaryClass = {astro-ph.IM},
       adsurl = {https://ui.adsabs.harvard.edu/abs/2016AJ....151...36L}
}

@ARTICLE{2019PASP..131l4504M,
       author = {{Meisner}, A.~M. and {Lang}, D. and {Schlafly}, E.~F. and {Schlegel}, D.~J.},
        title = "{unWISE Coadds: The Five-year Data Set}",
      journal = {\pasp},
         year = 2019,
        month = dec,
       volume = {131},
       number = {1006},
        pages = {124504},
          doi = {10.1088/1538-3873/ab3df4},
archivePrefix = {arXiv},
       eprint = {1909.05444},
 primaryClass = {astro-ph.IM},
       adsurl = {https://ui.adsabs.harvard.edu/abs/2019PASP..131l4504M}
}

@ARTICLE{2021ApJ...909...27H,
       author = {{Huang}, X. and {Storfer}, C. and {Gu}, A. and {Ravi}, V. and {Pilon}, A. and {Sheu}, W. and {Venguswamy}, R. and {Banka}, S. and {Dey}, A. and {Landriau}, M. and {Lang}, D. and {Meisner}, A. and {Moustakas}, J. and {Myers}, A.~D. and {Sajith}, R. and {Schlafly}, E.~F. and {Schlegel}, D.~J.},
        title = "{Discovering New Strong Gravitational Lenses in the DESI Legacy Imaging Surveys}",
      journal = {\apj},
         year = 2021,
        month = mar,
       volume = {909},
       number = {1},
          eid = {27},
        pages = {27},
          doi = {10.3847/1538-4357/abd62b},
archivePrefix = {arXiv},
       eprint = {2005.04730},
 primaryClass = {astro-ph.IM},
       adsurl = {https://ui.adsabs.harvard.edu/abs/2021ApJ...909...27H}
}

@ARTICLE{2010ARA&A..48...87T,
       author = {{Treu}, Tommaso},
        title = "{Strong Lensing by Galaxies}",
      journal = {\araa},
         year = 2010,
        month = sep,
       volume = {48},
        pages = {87-125},
          doi = {10.1146/annurev-astro-081309-130924},
archivePrefix = {arXiv},
       eprint = {1003.5567},
 primaryClass = {astro-ph.CO},
       adsurl = {https://ui.adsabs.harvard.edu/abs/2010ARA&A..48...87T}
}

@ARTICLE{2020Sci...369.1347M,
       author = {{Meneghetti}, Massimo and {Davoli}, Guido and {Bergamini}, Pietro and {Rosati}, Piero and {Natarajan}, Priyamvada and {Giocoli}, Carlo and {Caminha}, Gabriel B. and {Metcalf}, R. Benton and {Rasia}, Elena and {Borgani}, Stefano and {Calura}, Francesco and {Grillo}, Claudio and {Mercurio}, Amata and {Vanzella}, Eros},
        title = "{An excess of small-scale gravitational lenses observed in galaxy clusters}",
      journal = {Science},
         year = 2020,
        month = sep,
       volume = {369},
       number = {6509},
        pages = {1347-1351},
          doi = {10.1126/science.aax5164},
archivePrefix = {arXiv},
       eprint = {2009.04471},
 primaryClass = {astro-ph.GA},
       adsurl = {https://ui.adsabs.harvard.edu/abs/2020Sci...369.1347M}
}

@ARTICLE{2015ApJ...800...38G,
       author = {{Grillo}, C. and {Suyu}, S.~H. and {Rosati}, P. and {Mercurio}, A. and {Balestra}, I. and {Munari}, E. and {Nonino}, M. and {Caminha}, G.~B. and {Lombardi}, M. and {De Lucia}, G. and {Borgani}, S. and {Gobat}, R. and {Biviano}, A. and {Girardi}, M. and {Umetsu}, K. and {Coe}, D. and {Koekemoer}, A.~M. and {Postman}, M. and {Zitrin}, A. and {Halkola}, A. and {Broadhurst}, T. and {Sartoris}, B. and {Presotto}, V. and {Annunziatella}, M. and {Maier}, C. and {Fritz}, A. and {Vanzella}, E. and {Frye}, B.},
        title = "{CLASH-VLT: Insights on the Mass Substructures in the Frontier Fields Cluster MACS J0416.1-2403 through Accurate Strong Lens Modeling}",
      journal = {\apj},
         year = 2015,
        month = feb,
       volume = {800},
       number = {1},
          eid = {38},
        pages = {38},
          doi = {10.1088/0004-637X/800/1/38},
archivePrefix = {arXiv},
       eprint = {1407.7866},
 primaryClass = {astro-ph.CO},
       adsurl = {https://ui.adsabs.harvard.edu/abs/2015ApJ...800...38G}
}

@ARTICLE{1964MNRAS.128..307R,
       author = {{Refsdal}, S.},
        title = "{On the possibility of determining Hubble's parameter and the masses of galaxies from the gravitational lens effect}",
      journal = {\mnras},
         year = 1964,
        month = jan,
       volume = {128},
        pages = {307},
          doi = {10.1093/mnras/128.4.307},
       adsurl = {https://ui.adsabs.harvard.edu/abs/1964MNRAS.128..307R}
}

@ARTICLE{2017MNRAS.468.2590S,
       author = {{Suyu}, S.~H. and {Bonvin}, V. and {Courbin}, F. and {Fassnacht}, C.~D. and {Rusu}, C.~E. and {Sluse}, D. and {Treu}, T. and {Wong}, K.~C. and {Auger}, M.~W. and {Ding}, X. and {Hilbert}, S. and {Marshall}, P.~J. and {Rumbaugh}, N. and {Sonnenfeld}, A. and {Tewes}, M. and {Tihhonova}, O. and {Agnello}, A. and {Blandford}, R.~D. and {Chen}, G.~C. -F. and {Collett}, T. and {Koopmans}, L.~V.~E. and {Liao}, K. and {Meylan}, G. and {Spiniello}, C.},
        title = "{H0LiCOW - I. H$_{0}$ Lenses in COSMOGRAIL's Wellspring: program overview}",
      journal = {\mnras},
         year = 2017,
        month = jul,
       volume = {468},
       number = {3},
        pages = {2590-2604},
          doi = {10.1093/mnras/stx483},
archivePrefix = {arXiv},
       eprint = {1607.00017},
 primaryClass = {astro-ph.CO},
       adsurl = {https://ui.adsabs.harvard.edu/abs/2017MNRAS.468.2590S}
}

@ARTICLE{2020MNRAS.498.1420W,
       author = {{Wong}, Kenneth C. and {Suyu}, Sherry H. and {Chen}, Geoff C. -F. and {Rusu}, Cristian E. and {Millon}, Martin and {Sluse}, Dominique and {Bonvin}, Vivien and {Fassnacht}, Christopher D. and {Taubenberger}, Stefan and {Auger}, Matthew W. and {Birrer}, Simon and {Chan}, James H.~H. and {Courbin}, Frederic and {Hilbert}, Stefan and {Tihhonova}, Olga and {Treu}, Tommaso and {Agnello}, Adriano and {Ding}, Xuheng and {Jee}, Inh and {Komatsu}, Eiichiro and {Shajib}, Anowar J. and {Sonnenfeld}, Alessandro and {Blandford}, Roger D. and {Koopmans}, L{\'e}on V.~E. and {Marshall}, Philip J. and {Meylan}, Georges},
        title = "{H0LiCOW - XIII. A 2.4 per cent measurement of H$_{0}$ from lensed quasars: 5.3{\ensuremath{\sigma}} tension between early- and late-Universe probes}",
      journal = {\mnras},
         year = 2020,
        month = oct,
       volume = {498},
       number = {1},
        pages = {1420-1439},
          doi = {10.1093/mnras/stz3094},
archivePrefix = {arXiv},
       eprint = {1907.04869},
 primaryClass = {astro-ph.CO},
       adsurl = {https://ui.adsabs.harvard.edu/abs/2020MNRAS.498.1420W}
}

@ARTICLE{2024A&A...687A..81P,
       author = {{Palencia}, J.~M. and {Diego}, J.~M. and {Kavanagh}, B.~J. and {Mart{\'\i}nez-Arrizabalaga}, J.},
        title = "{Statistics of magnification for extremely lensed high redshift stars}",
      journal = {\aap},
         year = 2024,
        month = jul,
       volume = {687},
          eid = {A81},
        pages = {A81},
          doi = {10.1051/0004-6361/202347492},
archivePrefix = {arXiv},
       eprint = {2307.09505},
 primaryClass = {astro-ph.CO},
       adsurl = {https://ui.adsabs.harvard.edu/abs/2024A&A...687A..81P}
}

@ARTICLE{2018NatAs...2..334K,
       author = {{Kelly}, Patrick L. and {Diego}, Jose M. and {Rodney}, Steven and {Kaiser}, Nick and {Broadhurst}, Tom and {Zitrin}, Adi and {Treu}, Tommaso and {P{\'e}rez-Gonz{\'a}lez}, Pablo G. and {Morishita}, Takahiro and {Jauzac}, Mathilde and {Selsing}, Jonatan and {Oguri}, Masamune and {Pueyo}, Laurent and {Ross}, Timothy W. and {Filippenko}, Alexei V. and {Smith}, Nathan and {Hjorth}, Jens and {Cenko}, S. Bradley and {Wang}, Xin and {Howell}, D. Andrew and {Richard}, Johan and {Frye}, Brenda L. and {Jha}, Saurabh W. and {Foley}, Ryan J. and {Norman}, Colin and {Bradac}, Marusa and {Zheng}, Weikang and {Brammer}, Gabriel and {Benito}, Alberto Molino and {Cava}, Antonio and {Christensen}, Lise and {de Mink}, Selma E. and {Graur}, Or and {Grillo}, Claudio and {Kawamata}, Ryota and {Kneib}, Jean-Paul and {Matheson}, Thomas and {McCully}, Curtis and {Nonino}, Mario and {P{\'e}rez-Fournon}, Ismael and {Riess}, Adam G. and {Rosati}, Piero and {Schmidt}, Kasper Borello and {Sharon}, Keren and {Weiner}, Benjamin J.},
        title = "{Extreme magnification of an individual star at redshift 1.5 by a galaxy-cluster lens}",
      journal = {Nature Astronomy},
         year = 2018,
        month = apr,
       volume = {2},
        pages = {334-342},
          doi = {10.1038/s41550-018-0430-3},
archivePrefix = {arXiv},
       eprint = {1706.10279},
 primaryClass = {astro-ph.GA},
       adsurl = {https://ui.adsabs.harvard.edu/abs/2018NatAs...2..334K}
}

@ARTICLE{2022Natur.603..815W,
       author = {{Welch}, Brian and {Coe}, Dan and {Diego}, Jose M. and {Zitrin}, Adi and {Zackrisson}, Erik and {Dimauro}, Paola and {Jim{\'e}nez-Teja}, Yolanda and {Kelly}, Patrick and {Mahler}, Guillaume and {Oguri}, Masamune and {Timmes}, F.~X. and {Windhorst}, Rogier and {Florian}, Michael and {de Mink}, S.~E. and {Avila}, Roberto J. and {Anderson}, Jay and {Bradley}, Larry and {Sharon}, Keren and {Vikaeus}, Anton and {McCandliss}, Stephan and {Brada{\v{c}}}, Maru{\v{s}}a and {Rigby}, Jane and {Frye}, Brenda and {Toft}, Sune and {Strait}, Victoria and {Trenti}, Michele and {Sharma}, Soniya and {Andrade-Santos}, Felipe and {Broadhurst}, Tom},
        title = "{A highly magnified star at redshift 6.2}",
      journal = {\nat},
         year = 2022,
        month = mar,
       volume = {603},
       number = {7903},
        pages = {815-818},
          doi = {10.1038/s41586-022-04449-y},
archivePrefix = {arXiv},
       eprint = {2209.14866},
 primaryClass = {astro-ph.GA},
       adsurl = {https://ui.adsabs.harvard.edu/abs/2022Natur.603..815W}
}

@ARTICLE{1984Natur.310..112N,
       author = {{Narayan}, R. and {Blandford}, R. and {Nityananda}, R.},
        title = "{Multiple imaging of quasars by galaxies and clusters}",
      journal = {\nat},
         year = 1984,
        month = jul,
       volume = {310},
       number = {5973},
        pages = {112-115},
          doi = {10.1038/310112a0},
       adsurl = {https://ui.adsabs.harvard.edu/abs/1984Natur.310..112N}
}

@ARTICLE{2012ApJ...757...22C,
       author = {{Coe}, Dan and {Umetsu}, Keiichi and {Zitrin}, Adi and {Donahue}, Megan and {Medezinski}, Elinor and {Postman}, Marc and {Carrasco}, Mauricio and {Anguita}, Timo and {Geller}, Margaret J. and {Rines}, Kenneth J. and {Diaferio}, Antonaldo and {Kurtz}, Michael J. and {Bradley}, Larry and {Koekemoer}, Anton and {Zheng}, Wei and {Nonino}, Mario and {Molino}, Alberto and {Mahdavi}, Andisheh and {Lemze}, Doron and {Infante}, Leopoldo and {Ogaz}, Sara and {Melchior}, Peter and {Host}, Ole and {Ford}, Holland and {Grillo}, Claudio and {Rosati}, Piero and {Jim{\'e}nez-Teja}, Yolanda and {Moustakas}, John and {Broadhurst}, Tom and {Ascaso}, Bego{\~n}a and {Lahav}, Ofer and {Bartelmann}, Matthias and {Ben{\'\i}tez}, Narciso and {Bouwens}, Rychard and {Graur}, Or and {Graves}, Genevieve and {Jha}, Saurabh and {Jouvel}, Stephanie and {Kelson}, Daniel and {Moustakas}, Leonidas and {Maoz}, Dan and {Meneghetti}, Massimo and {Merten}, Julian and {Riess}, Adam and {Rodney}, Steve and {Seitz}, Stella},
        title = "{CLASH: Precise New Constraints on the Mass Profile of the Galaxy Cluster A2261}",
      journal = {\apj},
         year = 2012,
        month = sep,
       volume = {757},
       number = {1},
          eid = {22},
        pages = {22},
          doi = {10.1088/0004-637X/757/1/22},
archivePrefix = {arXiv},
       eprint = {1201.1616},
 primaryClass = {astro-ph.CO},
       adsurl = {https://ui.adsabs.harvard.edu/abs/2012ApJ...757...22C}
}

@ARTICLE{2010Sci...329..924J,
       author = {{Jullo}, E. and {Natarajan}, P. and {Kneib}, J. -P. and {D'Aloisio}, A. and {Limousin}, M. and {Richard}, J. and {Schimd}, C.},
        title = "{Cosmological constraints from strong gravitational lensing in clusters of galaxies.}",
      journal = {Science},
         year = 2010,
        month = aug,
       volume = {329},
        pages = {924-927},
          doi = {10.1126/science.1185759},
archivePrefix = {arXiv},
       eprint = {1008.4802},
 primaryClass = {astro-ph.CO},
       adsurl = {https://ui.adsabs.harvard.edu/abs/2010Sci...329..924J}
}

@ARTICLE{2013ApJ...762...32C,
       author = {{Coe}, Dan and {Zitrin}, Adi and {Carrasco}, Mauricio and {Shu}, Xinwen and {Zheng}, Wei and {Postman}, Marc and {Bradley}, Larry and {Koekemoer}, Anton and {Bouwens}, Rychard and {Broadhurst}, Tom and {Monna}, Anna and {Host}, Ole and {Moustakas}, Leonidas A. and {Ford}, Holland and {Moustakas}, John and {van der Wel}, Arjen and {Donahue}, Megan and {Rodney}, Steven A. and {Ben{\'\i}tez}, Narciso and {Jouvel}, Stephanie and {Seitz}, Stella and {Kelson}, Daniel D. and {Rosati}, Piero},
        title = "{CLASH: Three Strongly Lensed Images of a Candidate z {\ensuremath{\approx}} 11 Galaxy}",
      journal = {\apj},
         year = 2013,
        month = jan,
       volume = {762},
       number = {1},
          eid = {32},
        pages = {32},
          doi = {10.1088/0004-637X/762/1/32},
archivePrefix = {arXiv},
       eprint = {1211.3663},
 primaryClass = {astro-ph.CO},
       adsurl = {https://ui.adsabs.harvard.edu/abs/2013ApJ...762...32C}
}

@ARTICLE{2015ApJ...806....4M,
       author = {{Merten}, J. and {Meneghetti}, M. and {Postman}, M. and {Umetsu}, K. and {Zitrin}, A. and {Medezinski}, E. and {Nonino}, M. and {Koekemoer}, A. and {Melchior}, P. and {Gruen}, D. and {Moustakas}, L.~A. and {Bartelmann}, M. and {Host}, O. and {Donahue}, M. and {Coe}, D. and {Molino}, A. and {Jouvel}, S. and {Monna}, A. and {Seitz}, S. and {Czakon}, N. and {Lemze}, D. and {Sayers}, J. and {Balestra}, I. and {Rosati}, P. and {Ben{\'\i}tez}, N. and {Biviano}, A. and {Bouwens}, R. and {Bradley}, L. and {Broadhurst}, T. and {Carrasco}, M. and {Ford}, H. and {Grillo}, C. and {Infante}, L. and {Kelson}, D. and {Lahav}, O. and {Massey}, R. and {Moustakas}, J. and {Rasia}, E. and {Rhodes}, J. and {Vega}, J. and {Zheng}, W.},
        title = "{CLASH: The Concentration-Mass Relation of Galaxy Clusters}",
      journal = {\apj},
         year = 2015,
        month = jun,
       volume = {806},
       number = {1},
          eid = {4},
        pages = {4},
          doi = {10.1088/0004-637X/806/1/4},
archivePrefix = {arXiv},
       eprint = {1404.1376},
 primaryClass = {astro-ph.CO},
       adsurl = {https://ui.adsabs.harvard.edu/abs/2015ApJ...806....4M}
}

@ARTICLE{2019AJ....157..168D,
       author = {{Dey}, Arjun and {Schlegel}, David J. and {Lang}, Dustin and {Blum}, Robert and {Burleigh}, Kaylan and {Fan}, Xiaohui and {Findlay}, Joseph R. and {Finkbeiner}, Doug and {Herrera}, David and {Juneau}, St{\'e}phanie and {Landriau}, Martin and {Levi}, Michael and {McGreer}, Ian and {Meisner}, Aaron and {Myers}, Adam D. and {Moustakas}, John and {Nugent}, Peter and {Patej}, Anna and {Schlafly}, Edward F. and {Walker}, Alistair R. and {Valdes}, Francisco and {Weaver}, Benjamin A. and {Y{\`e}che}, Christophe and {Zou}, Hu and {Zhou}, Xu and {Abareshi}, Behzad and {Abbott}, T.~M.~C. and {Abolfathi}, Bela and {Aguilera}, C. and {Alam}, Shadab and {Allen}, Lori and {Alvarez}, A. and {Annis}, James and {Ansarinejad}, Behzad and {Aubert}, Marie and {Beechert}, Jacqueline and {Bell}, Eric F. and {BenZvi}, Segev Y. and {Beutler}, Florian and {Bielby}, Richard M. and {Bolton}, Adam S. and {Brice{\~n}o}, C{\'e}sar and {Buckley-Geer}, Elizabeth J. and {Butler}, Karen and {Calamida}, Annalisa and {Carlberg}, Raymond G. and {Carter}, Paul and {Casas}, Ricard and {Castander}, Francisco J. and {Choi}, Yumi and {Comparat}, Johan and {Cukanovaite}, Elena and {Delubac}, Timoth{\'e}e and {DeVries}, Kaitlin and {Dey}, Sharmila and {Dhungana}, Govinda and {Dickinson}, Mark and {Ding}, Zhejie and {Donaldson}, John B. and {Duan}, Yutong and {Duckworth}, Christopher J. and {Eftekharzadeh}, Sarah and {Eisenstein}, Daniel J. and {Etourneau}, Thomas and {Fagrelius}, Parker A. and {Farihi}, Jay and {Fitzpatrick}, Mike and {Font-Ribera}, Andreu and {Fulmer}, Leah and {G{\"a}nsicke}, Boris T. and {Gaztanaga}, Enrique and {George}, Koshy and {Gerdes}, David W. and {Gontcho}, Satya Gontcho A. and {Gorgoni}, Claudio and {Green}, Gregory and {Guy}, Julien and {Harmer}, Diane and {Hernandez}, M. and {Honscheid}, Klaus and {Huang}, Lijuan Wendy and {James}, David J. and {Jannuzi}, Buell T. and {Jiang}, Linhua and {Joyce}, Richard and {Karcher}, Armin and {Karkar}, Sonia and {Kehoe}, Robert and {Kneib}, Jean-Paul and {Kueter-Young}, Andrea and {Lan}, Ting-Wen and {Lauer}, Tod R. and {Le Guillou}, Laurent and {Le Van Suu}, Auguste and {Lee}, Jae Hyeon and {Lesser}, Michael and {Perreault Levasseur}, Laurence and {Li}, Ting S. and {Mann}, Justin L. and {Marshall}, Robert and {Mart{\'\i}nez-V{\'a}zquez}, C.~E. and {Martini}, Paul and {du Mas des Bourboux}, H{\'e}lion and {McManus}, Sean and {Meier}, Tobias Gabriel and {M{\'e}nard}, Brice and {Metcalfe}, Nigel and {Mu{\~n}oz-Guti{\'e}rrez}, Andrea and {Najita}, Joan and {Napier}, Kevin and {Narayan}, Gautham and {Newman}, Jeffrey A. and {Nie}, Jundan and {Nord}, Brian and {Norman}, Dara J. and {Olsen}, Knut A.~G. and {Paat}, Anthony and {Palanque-Delabrouille}, Nathalie and {Peng}, Xiyan and {Poppett}, Claire L. and {Poremba}, Megan R. and {Prakash}, Abhishek and {Rabinowitz}, David and {Raichoor}, Anand and {Rezaie}, Mehdi and {Robertson}, A.~N. and {Roe}, Natalie A. and {Ross}, Ashley J. and {Ross}, Nicholas P. and {Rudnick}, Gregory and {Safonova}, Sasha and {Saha}, Abhijit and {S{\'a}nchez}, F. Javier and {Savary}, Elodie and {Schweiker}, Heidi and {Scott}, Adam and {Seo}, Hee-Jong and {Shan}, Huanyuan and {Silva}, David R. and {Slepian}, Zachary and {Soto}, Christian and {Sprayberry}, David and {Staten}, Ryan and {Stillman}, Coley M. and {Stupak}, Robert J. and {Summers}, David L. and {Sien Tie}, Suk and {Tirado}, H. and {Vargas-Maga{\~n}a}, Mariana and {Vivas}, A. Katherina and {Wechsler}, Risa H. and {Williams}, Doug and {Yang}, Jinyi and {Yang}, Qian and {Yapici}, Tolga and {Zaritsky}, Dennis and {Zenteno}, A. and {Zhang}, Kai and {Zhang}, Tianmeng and {Zhou}, Rongpu and {Zhou}, Zhimin},
        title = "{Overview of the DESI Legacy Imaging Surveys}",
      journal = {\aj},
         year = 2019,
        month = may,
       volume = {157},
       number = {5},
          eid = {168},
        pages = {168},
          doi = {10.3847/1538-3881/ab089d},
archivePrefix = {arXiv},
       eprint = {1804.08657},
 primaryClass = {astro-ph.IM},
       adsurl = {https://ui.adsabs.harvard.edu/abs/2019AJ....157..168D}
}

@ARTICLE{2019ApJ...873..111I,
       author = {{Ivezi{\'c}}, {\v{Z}}eljko and {Kahn}, Steven M. and {Tyson}, J. Anthony and {Abel}, Bob and {Acosta}, Emily and {Allsman}, Robyn and {Alonso}, David and {AlSayyad}, Yusra and {Anderson}, Scott F. and {Andrew}, John and {Angel}, James Roger P. and {Angeli}, George Z. and {Ansari}, Reza and {Antilogus}, Pierre and {Araujo}, Constanza and {Armstrong}, Robert and {Arndt}, Kirk T. and {Astier}, Pierre and {Aubourg}, {\'E}ric and {Auza}, Nicole and {Axelrod}, Tim S. and {Bard}, Deborah J. and {Barr}, Jeff D. and {Barrau}, Aurelian and {Bartlett}, James G. and {Bauer}, Amanda E. and {Bauman}, Brian J. and {Baumont}, Sylvain and {Bechtol}, Ellen and {Bechtol}, Keith and {Becker}, Andrew C. and {Becla}, Jacek and {Beldica}, Cristina and {Bellavia}, Steve and {Bianco}, Federica B. and {Biswas}, Rahul and {Blanc}, Guillaume and {Blazek}, Jonathan and {Blandford}, Roger D. and {Bloom}, Josh S. and {Bogart}, Joanne and {Bond}, Tim W. and {Booth}, Michael T. and {Borgland}, Anders W. and {Borne}, Kirk and {Bosch}, James F. and {Boutigny}, Dominique and {Brackett}, Craig A. and {Bradshaw}, Andrew and {Brandt}, William Nielsen and {Brown}, Michael E. and {Bullock}, James S. and {Burchat}, Patricia and {Burke}, David L. and {Cagnoli}, Gianpietro and {Calabrese}, Daniel and {Callahan}, Shawn and {Callen}, Alice L. and {Carlin}, Jeffrey L. and {Carlson}, Erin L. and {Chandrasekharan}, Srinivasan and {Charles-Emerson}, Glenaver and {Chesley}, Steve and {Cheu}, Elliott C. and {Chiang}, Hsin-Fang and {Chiang}, James and {Chirino}, Carol and {Chow}, Derek and {Ciardi}, David R. and {Claver}, Charles F. and {Cohen-Tanugi}, Johann and {Cockrum}, Joseph J. and {Coles}, Rebecca and {Connolly}, Andrew J. and {Cook}, Kem H. and {Cooray}, Asantha and {Covey}, Kevin R. and {Cribbs}, Chris and {Cui}, Wei and {Cutri}, Roc and {Daly}, Philip N. and {Daniel}, Scott F. and {Daruich}, Felipe and {Daubard}, Guillaume and {Daues}, Greg and {Dawson}, William and {Delgado}, Francisco and {Dellapenna}, Alfred and {de Peyster}, Robert and {de Val-Borro}, Miguel and {Digel}, Seth W. and {Doherty}, Peter and {Dubois}, Richard and {Dubois-Felsmann}, Gregory P. and {Durech}, Josef and {Economou}, Frossie and {Eifler}, Tim and {Eracleous}, Michael and {Emmons}, Benjamin L. and {Fausti Neto}, Angelo and {Ferguson}, Henry and {Figueroa}, Enrique and {Fisher-Levine}, Merlin and {Focke}, Warren and {Foss}, Michael D. and {Frank}, James and {Freemon}, Michael D. and {Gangler}, Emmanuel and {Gawiser}, Eric and {Geary}, John C. and {Gee}, Perry and {Geha}, Marla and {Gessner}, Charles J.~B. and {Gibson}, Robert R. and {Gilmore}, D. Kirk and {Glanzman}, Thomas and {Glick}, William and {Goldina}, Tatiana and {Goldstein}, Daniel A. and {Goodenow}, Iain and {Graham}, Melissa L. and {Gressler}, William J. and {Gris}, Philippe and {Guy}, Leanne P. and {Guyonnet}, Augustin and {Haller}, Gunther and {Harris}, Ron and {Hascall}, Patrick A. and {Haupt}, Justine and {Hernandez}, Fabio and {Herrmann}, Sven and {Hileman}, Edward and {Hoblitt}, Joshua and {Hodgson}, John A. and {Hogan}, Craig and {Howard}, James D. and {Huang}, Dajun and {Huffer}, Michael E. and {Ingraham}, Patrick and {Innes}, Walter R. and {Jacoby}, Suzanne H. and {Jain}, Bhuvnesh and {Jammes}, Fabrice and {Jee}, M. James and {Jenness}, Tim and {Jernigan}, Garrett and {Jevremovi{\'c}}, Darko and {Johns}, Kenneth and {Johnson}, Anthony S. and {Johnson}, Margaret W.~G. and {Jones}, R. Lynne and {Juramy-Gilles}, Claire and {Juri{\'c}}, Mario and {Kalirai}, Jason S. and {Kallivayalil}, Nitya J. and {Kalmbach}, Bryce and {Kantor}, Jeffrey P. and {Karst}, Pierre and {Kasliwal}, Mansi M. and {Kelly}, Heather and {Kessler}, Richard and {Kinnison}, Veronica and {Kirkby}, David and {Knox}, Lloyd and {Kotov}, Ivan V. and {Krabbendam}, Victor L. and {Krughoff}, K. Simon and {Kub{\'a}nek}, Petr and {Kuczewski}, John and {Kulkarni}, Shri and {Ku}, John and {Kurita}, Nadine R. and {Lage}, Craig S. and {Lambert}, Ron and {Lange}, Travis and {Langton}, J. Brian and {Le Guillou}, Laurent and {Levine}, Deborah and {Liang}, Ming and {Lim}, Kian-Tat and {Lintott}, Chris J. and {Long}, Kevin E. and {Lopez}, Margaux and {Lotz}, Paul J. and {Lupton}, Robert H. and {Lust}, Nate B. and {MacArthur}, Lauren A. and {Mahabal}, Ashish and {Mandelbaum}, Rachel and {Markiewicz}, Thomas W. and {Marsh}, Darren S. and {Marshall}, Philip J. and {Marshall}, Stuart and {May}, Morgan and {McKercher}, Robert and {McQueen}, Michelle and {Meyers}, Joshua and {Migliore}, Myriam and {Miller}, Michelle and {Mills}, David J.},
        title = "{LSST: From Science Drivers to Reference Design and Anticipated Data Products}",
      journal = {\apj},
         year = 2019,
        month = mar,
       volume = {873},
       number = {2},
          eid = {111},
        pages = {111},
          doi = {10.3847/1538-4357/ab042c},
archivePrefix = {arXiv},
       eprint = {0805.2366},
 primaryClass = {astro-ph},
       adsurl = {https://ui.adsabs.harvard.edu/abs/2019ApJ...873..111I}
}

@ARTICLE{2025arXiv250315324E,
       author = {{Euclid Collaboration} and {Walmsley}, M. and {Holloway}, P. and {Lines}, N.~E.~P. and {Rojas}, K. and {Collett}, T.~E. and {Verma}, A. and {Li}, T. and {Nightingale}, J.~W. and {Despali}, G. and {Schuldt}, S. and {Gavazzi}, R. and {Melo}, A. and {Metcalf}, R.~B. and {Andika}, I.~T. and {Leuzzi}, L. and {Manj{\'o}n-Garc{\'\i}a}, A. and {Pearce-Casey}, R. and {Vincken}, S.~H. and {Wilde}, J. and {Busillo}, V. and {Tortora}, C. and {Acevedo Barroso}, J.~A. and {Dole}, H. and {Ecker}, L.~R. and {Pearson}, J. and {Marshall}, P.~J. and {More}, A. and {Saifollahi}, T. and {Gracia-Carpio}, J. and {Baeten}, E. and {Cornen}, C. and {Johnson}, L.~C. and {Macmillan}, C. and {Kruk}, S. and {Remmelgas}, K.~A. and {Cl{\'e}ment}, B. and {Degaudenzi}, H. and {Courbin}, F. and {Bovy}, J. and {Casas}, S. and {Dannerbauer}, H. and {Diego}, J.~M. and {Finner}, K. and {Galan}, A. and {Giocoli}, C. and {Hogg}, N.~B. and {Jahnke}, K. and {Katona}, J. and {Kov{\'a}cs}, A. and {De Leo}, C. and {Mahler}, G. and {Millon}, M. and {Nagam}, B.~C. and {Nugent}, P. and {Sainz de Murieta}, A. and {O'Riordan}, C.~M. and {Sluse}, D. and {Sonnenfeld}, A. and {Spiniello}, C. and {Serjeant}, S. and {Thai}, T.~T. and {Ulivi}, L. and {Walth}, G.~L. and {Weisenbach}, L. and {Zumalacarregui}, M. and {Aghanim}, N. and {Altieri}, B. and {Amara}, A. and {Andreon}, S. and {Auricchio}, N. and {Aussel}, H. and {Baccigalupi}, C. and {Baldi}, M. and {Balestra}, A. and {Bardelli}, S. and {Battaglia}, P. and {Bernardeau}, F. and {Biviano}, A. and {Bonchi}, A. and {Bonino}, D. and {Branchini}, E. and {Brescia}, M. and {Brinchmann}, J. and {Camera}, S. and {Ca{\~n}as-Herrera}, G. and {Capobianco}, V. and {Carbone}, C. and {Cardone}, V.~F. and {Carretero}, J. and {Castander}, F.~J. and {Castellano}, M. and {Castignani}, G. and {Cavuoti}, S. and {Chambers}, K.~C. and {Cimatti}, A. and {Colodro-Conde}, C. and {Congedo}, G. and {Conselice}, C.~J. and {Conversi}, L. and {Copin}, Y. and {Corcione}, L. and {Courtois}, H.~M. and {Cropper}, M. and {Da Silva}, A. and {De Lucia}, G. and {Di Giorgio}, A.~M. and {Dolding}, C. and {Dubath}, F. and {Duncan}, C.~A.~J. and {Dupac}, X. and {Ealet}, A. and {Escoffier}, S. and {Fabricius}, M. and {Farina}, M. and {Farinelli}, R. and {Faustini}, F. and {Finelli}, F. and {Fotopoulou}, S. and {Frailis}, M. and {Franceschi}, E. and {Fumana}, M. and {Galeotta}, S. and {George}, K. and {Gillard}, W. and {Gillis}, B. and {G{\'o}mez-Alvarez}, P. and {Granett}, B.~R. and {Grazian}, A. and {Grupp}, F. and {Guzzo}, L. and {Gwyn}, S. and {Haugan}, S.~V.~H. and {Hoekstra}, H. and {Holmes}, W. and {Hook}, I.~M. and {Hormuth}, F. and {Hornstrup}, A. and {Hudelot}, P. and {Jhabvala}, M. and {Joachimi}, B. and {Keih{\"a}nen}, E. and {Kermiche}, S. and {Kiessling}, A. and {Kubik}, B. and {K{\"u}mmel}, M. and {Kunz}, M. and {Kurki-Suonio}, H. and {Lahav}, O. and {Le Boulc'h}, Q. and {Le Brun}, A.~M.~C. and {Le Mignant}, D. and {Ligori}, S. and {Lilje}, P.~B. and {Lindholm}, V. and {Lloro}, I. and {Mainetti}, G. and {Maino}, D. and {Maiorano}, E. and {Mansutti}, O. and {Marcin}, S. and {Marggraf}, O. and {Martinelli}, M. and {Martinet}, N. and {Marulli}, F. and {Massey}, R. and {Maurogordato}, S. and {McCracken}, H.~J. and {Medinaceli}, E. and {Mei}, S. and {Mellier}, Y. and {Meneghetti}, M. and {Merlin}, E. and {Meylan}, G. and {Mora}, A. and {Moresco}, M. and {Moscardini}, L. and {Nakajima}, R. and {Neissner}, C. and {Nichol}, R.~C. and {Niemi}, S. -M. and {Padilla}, C. and {Paltani}, S. and {Pasian}, F. and {Pedersen}, K. and {Percival}, W.~J. and {Pettorino}, V. and {Pires}, S. and {Polenta}, G. and {Poncet}, M. and {Popa}, L.~A. and {Pozzetti}, L. and {Raison}, F. and {Rebolo}, R. and {Renzi}, A. and {Rhodes}, J. and {Riccio}, G. and {Romelli}, E. and {Roncarelli}, M. and {Saglia}, R.},
        title = "{Euclid Quick Data Release (Q1): The Strong Lensing Discovery Engine A -- System overview and lens catalogue}",
      journal = {arXiv e-prints},
         year = 2025,
        month = mar,
          eid = {arXiv:2503.15324},
        pages = {arXiv:2503.15324},
          doi = {10.48550/arXiv.2503.15324},
archivePrefix = {arXiv},
       eprint = {2503.15324},
 primaryClass = {astro-ph.GA},
       adsurl = {https://ui.adsabs.harvard.edu/abs/2025arXiv250315324E}
}

@ARTICLE{2021ApJ...906..107K,
       author = {{Khullar}, Gourav and {Gozman}, Katya and {Lin}, Jason J. and {Martinez}, Michael N. and {Matthews Acu{\~n}a}, Owen S. and {Medina}, Elisabeth and {Merz}, Kaiya and {Sanchez}, Jorge A. and {Sisco}, Emily E. and {Kavin Stein}, Daniel J. and {Sukay}, Ezra O. and {Tavangar}, Kiyan and {Bayliss}, Matthew B. and {Bleem}, Lindsey E. and {Brownsberger}, Sasha and {Dahle}, H{\r{A}}kon and {Florian}, Michael K. and {Gladders}, Michael D. and {Mahler}, Guillaume and {Rigby}, Jane R. and {Sharon}, Keren and {Stark}, Antony A.},
        title = "{COOL-LAMPS. I. An Extraordinarily Bright Lensed Galaxy at Redshift 5.04}",
      journal = {\apj},
         year = 2021,
        month = jan,
       volume = {906},
       number = {2},
          eid = {107},
        pages = {107},
          doi = {10.3847/1538-4357/abcb86},
archivePrefix = {arXiv},
       eprint = {2011.06601},
 primaryClass = {astro-ph.GA},
       adsurl = {https://ui.adsabs.harvard.edu/abs/2021ApJ...906..107K}
}

@ARTICLE{2016ApJ...817...85X,
       author = {{Xu}, Bingxiao and {Postman}, Marc and {Meneghetti}, Massimo and {Seitz}, Stella and {Zitrin}, Adi and {Merten}, Julian and {Maoz}, Dani and {Frye}, Brenda and {Umetsu}, Keiichi and {Zheng}, Wei and {Bradley}, Larry and {Vega}, Jesus and {Koekemoer}, Anton},
        title = "{The Detection and Statistics of Giant Arcs behind CLASH Clusters}",
      journal = {\apj},
         year = 2016,
        month = feb,
       volume = {817},
       number = {2},
          eid = {85},
        pages = {85},
          doi = {10.3847/0004-637X/817/2/85},
archivePrefix = {arXiv},
       eprint = {1511.04002},
 primaryClass = {astro-ph.CO},
       adsurl = {https://ui.adsabs.harvard.edu/abs/2016ApJ...817...85X}
}

@ARTICLE{2015arXiv150203167I,
       author = {{Ioffe}, Sergey and {Szegedy}, Christian},
        title = "{Batch Normalization: Accelerating Deep Network Training by Reducing Internal Covariate Shift}",
      journal = {arXiv e-prints},
         year = 2015,
        month = feb,
          eid = {arXiv:1502.03167},
        pages = {arXiv:1502.03167},
          doi = {10.48550/arXiv.1502.03167},
archivePrefix = {arXiv},
       eprint = {1502.03167},
 primaryClass = {cs.LG},
       adsurl = {https://ui.adsabs.harvard.edu/abs/2015arXiv150203167I}
}

@ARTICLE{2021ApJ...923...16L,
       author = {{Li}, R. and {Napolitano}, N.~R. and {Spiniello}, C. and {Tortora}, C. and {Kuijken}, K. and {Koopmans}, L.~V.~E. and {Schneider}, P. and {Getman}, F. and {Xie}, L. and {Long}, L. and {Shu}, W. and {Vernardos}, G. and {Huang}, Z. and {Covone}, G. and {Dvornik}, A. and {Heymans}, C. and {Hildebrandt}, H. and {Radovich}, M. and {Wright}, A.~H.},
        title = "{High-quality Strong Lens Candidates in the Final Kilo-Degree Survey Footprint}",
      journal = {\apj},
         year = 2021,
        month = dec,
       volume = {923},
       number = {1},
          eid = {16},
        pages = {16},
          doi = {10.3847/1538-4357/ac2df0},
archivePrefix = {arXiv},
       eprint = {2110.01905},
 primaryClass = {astro-ph.GA},
       adsurl = {https://ui.adsabs.harvard.edu/abs/2021ApJ...923...16L}
}

@ARTICLE{2017MNRAS.472.1129P,
       author = {{Petrillo}, C.~E. and {Tortora}, C. and {Chatterjee}, S. and {Vernardos}, G. and {Koopmans}, L.~V.~E. and {Verdoes Kleijn}, G. and {Napolitano}, N.~R. and {Covone}, G. and {Schneider}, P. and {Grado}, A. and {McFarland}, J.},
        title = "{Finding strong gravitational lenses in the Kilo Degree Survey with Convolutional Neural Networks}",
      journal = {\mnras},
         year = 2017,
        month = nov,
       volume = {472},
       number = {1},
        pages = {1129-1150},
          doi = {10.1093/mnras/stx2052},
archivePrefix = {arXiv},
       eprint = {1702.07675},
 primaryClass = {astro-ph.GA},
       adsurl = {https://ui.adsabs.harvard.edu/abs/2017MNRAS.472.1129P}
}

@ARTICLE{2019ApJS..243...17J,
       author = {{Jacobs}, C. and {Collett}, T. and {Glazebrook}, K. and {Buckley-Geer}, E. and {Diehl}, H.~T. and {Lin}, H. and {McCarthy}, C. and {Qin}, A.~K. and {Odden}, C. and {Caso Escudero}, M. and {Dial}, P. and {Yung}, V.~J. and {Gaitsch}, S. and {Pellico}, A. and {Lindgren}, K.~A. and {Abbott}, T.~M.~C. and {Annis}, J. and {Avila}, S. and {Brooks}, D. and {Burke}, D.~L. and {Carnero Rosell}, A. and {Carrasco Kind}, M. and {Carretero}, J. and {da Costa}, L.~N. and {De Vicente}, J. and {Fosalba}, P. and {Frieman}, J. and {Garc{\'\i}a-Bellido}, J. and {Gaztanaga}, E. and {Goldstein}, D.~A. and {Gruen}, D. and {Gruendl}, R.~A. and {Gschwend}, J. and {Hollowood}, D.~L. and {Honscheid}, K. and {Hoyle}, B. and {James}, D.~J. and {Krause}, E. and {Kuropatkin}, N. and {Lahav}, O. and {Lima}, M. and {Maia}, M.~A.~G. and {Marshall}, J.~L. and {Miquel}, R. and {Plazas}, A.~A. and {Roodman}, A. and {Sanchez}, E. and {Scarpine}, V. and {Serrano}, S. and {Sevilla-Noarbe}, I. and {Smith}, M. and {Sobreira}, F. and {Suchyta}, E. and {Swanson}, M.~E.~C. and {Tarle}, G. and {Vikram}, V. and {Walker}, A.~R. and {Zhang}, Y. and {DES Collaboration}},
        title = "{An Extended Catalog of Galaxy-Galaxy Strong Gravitational Lenses Discovered in DES Using Convolutional Neural Networks}",
      journal = {\apjs},
         year = 2019,
        month = jul,
       volume = {243},
       number = {1},
          eid = {17},
        pages = {17},
          doi = {10.3847/1538-4365/ab26b6},
archivePrefix = {arXiv},
       eprint = {1905.10522},
 primaryClass = {astro-ph.GA},
       adsurl = {https://ui.adsabs.harvard.edu/abs/2019ApJS..243...17J}
}

@ARTICLE{2024MNRAS.535.1625J,
       author = {{Jaelani}, Anton T. and {More}, Anupreeta and {Wong}, Kenneth C. and {Inoue}, Kaiki T. and {Chao}, Dani C. -Y. and {Premadi}, Premana W. and {Ca{\~n}ameras}, Raoul},
        title = "{Survey of gravitationally lensed objects in HSC imaging (SuGOHI) - X. Strong lens finding in the HSC-SSP using convolutional neural networks}",
      journal = {\mnras},
         year = 2024,
        month = dec,
       volume = {535},
       number = {2},
        pages = {1625-1639},
          doi = {10.1093/mnras/stae2442},
archivePrefix = {arXiv},
       eprint = {2312.07333},
 primaryClass = {astro-ph.GA},
       adsurl = {https://ui.adsabs.harvard.edu/abs/2024MNRAS.535.1625J}
}

@misc{kingma2017adammethodstochasticoptimization,
      title={Adam: A Method for Stochastic Optimization}, 
      author={Diederik P. Kingma and Jimmy Ba},
      year={2017},
      eprint={1412.6980},
      archivePrefix={arXiv},
      primaryClass={cs.LG},
      url={https://arxiv.org/abs/1412.6980}, 
}

@misc{dosovitskiy2021imageworth16x16words,
      title={An Image is Worth 16x16 Words: Transformers for Image Recognition at Scale}, 
      author={Alexey Dosovitskiy and Lucas Beyer and Alexander Kolesnikov and Dirk Weissenborn and Xiaohua Zhai and Thomas Unterthiner and Mostafa Dehghani and Matthias Minderer and Georg Heigold and Sylvain Gelly and Jakob Uszkoreit and Neil Houlsby},
      year={2021},
      eprint={2010.11929},
      archivePrefix={arXiv},
      primaryClass={cs.CV},
      url={https://arxiv.org/abs/2010.11929}, 
}

@ARTICLE{2011arXiv1112.5745H,
       author = {{Houlsby}, Neil and {Husz{\'a}r}, Ferenc and {Ghahramani}, Zoubin and {Lengyel}, M{\'a}t{\'e}},
        title = "{Bayesian Active Learning for Classification and Preference Learning}",
      journal = {arXiv e-prints},
         year = 2011,
        month = dec,
          eid = {arXiv:1112.5745},
        pages = {arXiv:1112.5745},
          doi = {10.48550/arXiv.1112.5745},
archivePrefix = {arXiv},
       eprint = {1112.5745},
 primaryClass = {stat.ML},
       adsurl = {https://ui.adsabs.harvard.edu/abs/2011arXiv1112.5745H}
}

@ARTICLE{2022AJ....163..139Y,
       author = {{Yue}, Minghao and {Fan}, Xiaohui and {Yang}, Jinyi and {Wang}, Feige},
        title = "{A Mock Catalog of Gravitationally-lensed Quasars for the LSST Survey}",
      journal = {\aj},
         year = 2022,
        month = mar,
       volume = {163},
       number = {3},
          eid = {139},
        pages = {139},
          doi = {10.3847/1538-3881/ac4cb0},
archivePrefix = {arXiv},
       eprint = {2201.06761},
 primaryClass = {astro-ph.GA},
       adsurl = {https://ui.adsabs.harvard.edu/abs/2022AJ....163..139Y}
}

@article{Xu_2019,
   title={A study of stellar orbit fractions: simulated IllustrisTNG galaxies compared to CALIFA observations},
   volume={489},
   ISSN={1365-2966},
   url={http://dx.doi.org/10.1093/mnras/stz2164},
   DOI={10.1093/mnras/stz2164},
   number={1},
   journal={Monthly Notices of the Royal Astronomical Society},
   publisher={Oxford University Press (OUP)},
   author={Xu, Dandan and Zhu, Ling and Grand, Robert and Springel, Volker and Mao, Shude and van de Ven, Glenn and Lu, Shengdong and Wang, Yougang and Pillepich, Annalisa and Genel, Shy and Nelson, Dylan and Rodriguez-Gomez, Vicente and Pakmor, Rüdiger and Weinberger, Rainer and Marinacci, Federico and Vogelsberger, Mark and Torrey, Paul and Naiman, Jill and Hernquist, Lars},
   year={2019},
   month=aug, pages={842–854} }

@ARTICLE{2002A&A...390..821G,
       author = {{Golse}, G. and {Kneib}, J. -P.},
        title = "{Pseudo elliptical lensing mass model: Application to the NFW mass distribution}",
      journal = {\aap},
         year = 2002,
        month = aug,
       volume = {390},
        pages = {821-827},
          doi = {10.1051/0004-6361:20020639},
archivePrefix = {arXiv},
       eprint = {astro-ph/0112138},
 primaryClass = {astro-ph},
       adsurl = {https://ui.adsabs.harvard.edu/abs/2002A&A...390..821G}
}

@misc{desicollaboration2025datarelease1dark,
      title={Data Release 1 of the Dark Energy Spectroscopic Instrument}, 
      author={DESI Collaboration and M. Abdul-Karim and A. G. Adame and D. Aguado and J. Aguilar and S. Ahlen and S. Alam and G. Aldering and D. M. Alexander and R. Alfarsy and L. Allen and C. Allende Prieto and O. Alves and A. Anand and U. Andrade and E. Armengaud and S. Avila and A. Aviles and H. Awan and S. Bailey and A. Baleato Lizancos and O. Ballester and A. Bault and J. Bautista and S. BenZvi and L. Beraldo e Silva and J. R. Bermejo-Climent and F. Beutler and D. Bianchi and C. Blake and R. Blum and A. S. Bolton and M. Bonici and S. Brieden and A. Brodzeller and D. Brooks and E. Buckley-Geer and E. Burtin and R. Canning and A. Carnero Rosell and A. Carr and P. Carrilho and L. Casas and F. J. Castander and R. Cereskaite and J. L. Cervantes-Cota and E. Chaussidon and J. Chaves-Montero and S. Chen and X. Chen and T. Claybaugh and S. Cole and A. P. Cooper and M. -C. Cousinou and A. Cuceu and T. M. Davis and K. S. Dawson and R. de Belsunce and R. de la Cruz and A. de la Macorra and A. de Mattia and N. Deiosso and J. Della Costa and R. Demina and U. Demirbozan and J. DeRose and A. Dey and B. Dey and J. Ding and Z. Ding and P. Doel and K. Douglass and M. Dowicz and H. Ebina and J. Edelstein and D. J. Eisenstein and W. Elbers and N. Emas and S. Escoffier and P. Fagrelius and X. Fan and K. Fanning and V. A. Fawcett and E. Fernández-García and S. Ferraro and N. Findlay and A. Font-Ribera and J. E. Forero-Romero and D. Forero-Sánchez and C. S. Frenk and B. T. Gänsicke and L. Galbany and J. García-Bellido and C. Garcia-Quintero and L. H. Garrison and E. Gaztañaga and H. Gil-Marín and O. Y. Gnedin and S. Gontcho A Gontcho and A. X. Gonzalez-Morales and V. Gonzalez-Perez and C. Gordon and O. Graur and D. Green and D. Gruen and R. Gsponer and C. Guandalin and G. Gutierrez and J. Guy and C. Hahn and J. J. Han and J. Han and S. He and H. K. Herrera-Alcantar and K. Honscheid and J. Hou and C. Howlett and D. Huterer and V. Iršič and M. Ishak and A. Jacques and J. Jimenez and Y. P. Jing and B. Joachimi and S. Joudaki and R. Joyce and E. Jullo and S. Juneau and N. G. Karaçaylı and T. Karim and R. Kehoe and S. Kent and A. Khederlarian and D. Kirkby and T. Kisner and F. -S. Kitaura and N. Kizhuprakkat and H. Kong and S. E. Koposov and A. Kremin and A. Krolewski and O. Lahav and Y. Lai and C. Lamman and T. -W. Lan and M. Landriau and D. Lang and J. U. Lange and J. Lasker and J. M. Le Goff and L. Le Guillou and A. Leauthaud and M. E. Levi and S. Li and T. S. Li and K. Lodha and M. Lokken and Y. Luo and C. Magneville and M. Manera and C. J. Manser and D. Margala and P. Martini and M. Maus and J. McCullough and P. McDonald and G. E. Medina and L. Medina-Varela and A. Meisner and J. Mena-Fernández and A. Menegas and M. Mezcua and R. Miquel and P. Montero-Camacho and J. Moon and J. Moustakas and A. Muñoz-Gutiérrez and D. Muñoz-Santos and A. D. Myers and J. Myles and S. Nadathur and J. Najita and L. Napolitano and J. A. Newman and F. Nikakhtar and R. Nikutta and G. Niz and H. E. Noriega and N. Padmanabhan and E. Paillas and N. Palanque-Delabrouille and A. Palmese and J. Pan and Z. Pan and D. Parkinson and J. Peacock and W. J. Percival and A. Pérez-Fernández and I. Pérez-Ràfols and P. Peterson and J. Piat and M. M. Pieri and M. Pinon and C. Poppett and A. Porredon and F. Prada and R. Pucha and F. Qin and D. Rabinowitz and A. Raichoor and C. Ramírez-Pérez and S. Ramirez-Solano and M. Rashkovetskyi and C. Ravoux and A. H. Riley and A. Rocher and C. Rockosi and J. Rohlf and A. J. Ross and G. Rossi and R. Ruggeri and V. Ruhlmann-Kleider and C. G. Sabiu and K. Said and A. Saintonge and L. Samushia and E. Sanchez and N. Sanders and C. Saulder and E. F. Schlafly and D. Schlegel and D. Scholte and M. Schubnell and H. Seo and A. Shafieloo and R. Sharples and J. Silber and M. Siudek and A. Smith and D. Sprayberry and J. Suárez-Pérez and J. Swanson and T. Tan and G. Tarlé and P. Taylor and G. Thomas and R. Tojeiro and R. J. Turner and W. Turner and L. A. Ureña-López and R. Vaisakh and M. Valluri and M. Vargas-Magaña and L. Verde and M. Walther and B. Wang and M. S. Wang and W. Wang and B. A. Weaver and N. Weaverdyck and R. H. Wechsler and M. White and M. Wolfson and J. Yang and C. Yèche and S. Youles and J. Yu and S. Yuan and E. A. Zaborowski and P. Zarrouk and H. Zhang and C. Zhao and R. Zhao and Z. Zheng and R. Zhou and H. Zou and S. Zou and Y. Zu},
      year={2025},
      eprint={2503.14745},
      archivePrefix={arXiv},
      primaryClass={astro-ph.CO},
      url={https://arxiv.org/abs/2503.14745}, 
}

@ARTICLE{1958ApJS....3..211A,
       author = {{Abell}, George O.},
        title = "{The Distribution of Rich Clusters of Galaxies.}",
      journal = {\apjs},
         year = 1958,
        month = may,
       volume = {3},
        pages = {211},
          doi = {10.1086/190036},
       adsurl = {https://ui.adsabs.harvard.edu/abs/1958ApJS....3..211A}
}

@ARTICLE{2011A&ARv..19...47K,
       author = {{Kneib}, Jean-Paul and {Natarajan}, Priyamvada},
        title = "{Cluster lenses}",
      journal = {\aapr},
         year = 2011,
        month = nov,
       volume = {19},
          eid = {47},
        pages = {47},
          doi = {10.1007/s00159-011-0047-3},
archivePrefix = {arXiv},
       eprint = {1202.0185},
 primaryClass = {astro-ph.CO},
       adsurl = {https://ui.adsabs.harvard.edu/abs/2011A&ARv..19...47K}
}

@ARTICLE{2012ApJS..199...25P,
       author = {{Postman}, Marc and {Coe}, Dan and {Ben{\'\i}tez}, Narciso and {Bradley}, Larry and {Broadhurst}, Tom and {Donahue}, Megan and {Ford}, Holland and {Graur}, Or and {Graves}, Genevieve and {Jouvel}, Stephanie and {Koekemoer}, Anton and {Lemze}, Doron and {Medezinski}, Elinor and {Molino}, Alberto and {Moustakas}, Leonidas and {Ogaz}, Sara and {Riess}, Adam and {Rodney}, Steve and {Rosati}, Piero and {Umetsu}, Keiichi and {Zheng}, Wei and {Zitrin}, Adi and {Bartelmann}, Matthias and {Bouwens}, Rychard and {Czakon}, Nicole and {Golwala}, Sunil and {Host}, Ole and {Infante}, Leopoldo and {Jha}, Saurabh and {Jimenez-Teja}, Yolanda and {Kelson}, Daniel and {Lahav}, Ofer and {Lazkoz}, Ruth and {Maoz}, Dani and {McCully}, Curtis and {Melchior}, Peter and {Meneghetti}, Massimo and {Merten}, Julian and {Moustakas}, John and {Nonino}, Mario and {Patel}, Brandon and {Reg{\"o}s}, Enik{\"o} and {Sayers}, Jack and {Seitz}, Stella and {Van der Wel}, Arjen},
        title = "{The Cluster Lensing and Supernova Survey with Hubble: An Overview}",
      journal = {\apjs},
         year = 2012,
        month = apr,
       volume = {199},
       number = {2},
          eid = {25},
        pages = {25},
          doi = {10.1088/0067-0049/199/2/25},
archivePrefix = {arXiv},
       eprint = {1106.3328},
 primaryClass = {astro-ph.CO},
       adsurl = {https://ui.adsabs.harvard.edu/abs/2012ApJS..199...25P}
}

@ARTICLE{2010PASJ...62.1017O,
       author = {{Oguri}, Masamune},
        title = "{The Mass Distribution of SDSS J1004+4112 Revisited}",
      journal = {\pasj},
         year = 2010,
        month = aug,
       volume = {62},
        pages = {1017},
          doi = {10.1093/pasj/62.4.1017},
archivePrefix = {arXiv},
       eprint = {1005.3103},
 primaryClass = {astro-ph.CO},
       adsurl = {https://ui.adsabs.harvard.edu/abs/2010PASJ...62.1017O}
}

@ARTICLE{2004A&A...416..391L,
       author = {{Lenzen}, F. and {Schindler}, S. and {Scherzer}, O.},
        title = "{Automatic detection of arcs and arclets formed by gravitational lensing}",
      journal = {\aap},
         year = 2004,
        month = mar,
       volume = {416},
        pages = {391-401},
          doi = {10.1051/0004-6361:20034619},
archivePrefix = {arXiv},
       eprint = {astro-ph/0311554},
 primaryClass = {astro-ph},
       adsurl = {https://ui.adsabs.harvard.edu/abs/2004A&A...416..391L}
}

@ARTICLE{2005ApJ...633..768H,
       author = {{Horesh}, Assaf and {Ofek}, Eran O. and {Maoz}, Dan and {Bartelmann}, Matthias and {Meneghetti}, Massimo and {Rix}, Hans-Walter},
        title = "{The Lensed Arc Production Efficiency of Galaxy Clusters: A Comparison of Matched Observed and Simulated Samples}",
      journal = {\apj},
         year = 2005,
        month = nov,
       volume = {633},
       number = {2},
        pages = {768-780},
          doi = {10.1086/466519},
archivePrefix = {arXiv},
       eprint = {astro-ph/0507454},
 primaryClass = {astro-ph},
       adsurl = {https://ui.adsabs.harvard.edu/abs/2005ApJ...633..768H}
}

@ARTICLE{2006astro.ph..6757A,
       author = {{Alard}, C.},
        title = "{Automated detection of gravitational arcs}",
      journal = {arXiv e-prints},
         year = 2006,
        month = jun,
          eid = {astro-ph/0606757},
        pages = {astro-ph/0606757},
          doi = {10.48550/arXiv.astro-ph/0606757},
archivePrefix = {arXiv},
       eprint = {astro-ph/0606757},
 primaryClass = {astro-ph},
       adsurl = {https://ui.adsabs.harvard.edu/abs/2006astro.ph..6757A}
}

@ARTICLE{2007A&A...472..341S,
       author = {{Seidel}, G. and {Bartelmann}, M.},
        title = "{Arcfinder: an algorithm for the automatic detection of gravitational arcs}",
      journal = {\aap},
         year = 2007,
        month = sep,
       volume = {472},
       number = {1},
        pages = {341-352},
          doi = {10.1051/0004-6361:20066097},
archivePrefix = {arXiv},
       eprint = {astro-ph/0607547},
 primaryClass = {astro-ph},
       adsurl = {https://ui.adsabs.harvard.edu/abs/2007A&A...472..341S}
}

@ARTICLE{1996astro.ph..6001N,
       author = {{Narayan}, Ramesh and {Bartelmann}, Matthias},
        title = "{Lectures on Gravitational Lensing}",
      journal = {arXiv e-prints},
         year = 1996,
        month = jun,
          eid = {astro-ph/9606001},
        pages = {astro-ph/9606001},
          doi = {10.48550/arXiv.astro-ph/9606001},
archivePrefix = {arXiv},
       eprint = {astro-ph/9606001},
 primaryClass = {astro-ph},
       adsurl = {https://ui.adsabs.harvard.edu/abs/1996astro.ph..6001N}
}

@ARTICLE{1988MNRAS.231P..97N,
       author = {{Narayan}, Ramesh and {White}, Simon D.~M.},
        title = "{Gravitational lensing in a cold dark matter universe}",
      journal = {\mnras},
         year = 1988,
        month = apr,
       volume = {231},
        pages = {97p-103p},
          doi = {10.1093/mnras/231.1.97P},
       adsurl = {https://ui.adsabs.harvard.edu/abs/1988MNRAS.231P..97N}
}

@ARTICLE{1996ApJ...471..643K,
       author = {{Kneib}, J. -P. and {Ellis}, R.~S. and {Smail}, I. and {Couch}, W.~J. and {Sharples}, R.~M.},
        title = "{Hubble Space Telescope Observations of the Lensing Cluster Abell 2218}",
      journal = {\apj},
         year = 1996,
        month = nov,
       volume = {471},
        pages = {643},
          doi = {10.1086/177995},
archivePrefix = {arXiv},
       eprint = {astro-ph/9511015},
 primaryClass = {astro-ph},
       adsurl = {https://ui.adsabs.harvard.edu/abs/1996ApJ...471..643K}
}

@ARTICLE{1996ApJ...464...92G,
       author = {{Grogin}, Norman A. and {Narayan}, Ramesh},
        title = "{A New Model of the Gravitational Lens 0957+561 and a Limit on the Hubble Constant}",
      journal = {\apj},
         year = 1996,
        month = jun,
       volume = {464},
        pages = {92},
          doi = {10.1086/177302},
       adsurl = {https://ui.adsabs.harvard.edu/abs/1996ApJ...464...92G}
}

@ARTICLE{1993A&A...273..367K,
       author = {{Kneib}, J. -P. and {Mellier}, Y. and {Fort}, B. and {Mathez}, G.},
        title = "{The distribution of dark matter in distant cluster-lenses : modelling modelling A 370.}",
      journal = {\aap},
         year = 1993,
        month = jun,
       volume = {273},
        pages = {367},
       adsurl = {https://ui.adsabs.harvard.edu/abs/1993A&A...273..367K}
}
\bibliographystyle{aasjournal}

%\bibliography{ref}{}
%\bibliographystyle{abbrv}
%\end{document}

\end{document}